\documentclass[pre]{revtex4}

\usepackage{graphicx}
\usepackage{amssymb,amsmath}
\begin{document}
\title{Inhomogeneous percolation models for spreading phenomena in random graphs}
\author{Luca Dall'Asta}
\address{Laboratoire de Physique Th\'eorique (UMR du CNRS 8627), B\^atiment 210, Universit\'e de Paris-Sud, 91405 ORSAY Cedex (France)}
\email{luca.dallasta@th.u-psud.fr}

\begin{abstract}
Percolation theory has been largely used in the study of structural properties of complex networks such as the robustness, 
with remarkable results. Nevertheless, a purely topological description is not sufficient for a correct characterization 
of networks behaviour in relation with physical flows and spreading phenomena taking place on them. 
The functionality of real networks also depends on the ability of the nodes and the edges in bearing and handling loads of flows, energy, information and other physical quantities.   
We propose to study these properties introducing a process of inhomogeneous percolation, in which both the nodes and the edges spread out the flows with a given probability.  

Generating functions approach is exploited in order to get a generalization of the Molloy-Reed Criterion [M. Molloy and B. Reed, {\em Random Struct. and Alg.} {\bf 6}, 161-180 (1995)] for inhomogeneous joint site bond percolation in correlated random graphs. A series of simple assumptions allows the analysis of more realistic situations, for which a number of new results are presented. In particular, for the site percolation with inhomogeneous edge transmission, we obtain the explicit expressions of the percolation threshold for many interesting cases, that are analyzed by means of simple examples and numerical simulations. Some possible applications are debated. 
\end{abstract}

\maketitle

\section{INTRODUCTION}
\label{intro}

First introduced in the study of gelation processes in the $1940$s \cite{flory,hammersley}, percolation theory has been largely used in statistical mechanics in order to describe topological phase transitions in lattice systems, such as the emergence of global properties when some physical parameter (temperature, concentration, etc.) exceeds a critical value \cite{stauffer,essam,bunde12,kirkpatrick}. 
More precisely, a site percolation process can be sketched as follows. Each site of the lattice is occupied with a uniform probability $q$, called {\em occupation probability}, and empty with probability $1-q$. A cluster is a connected set of occupied sites. When the value $q$ of the occupation probability is larger than a critical value $q_c$, the {\em percolation threshold}, a cluster with size of the same order of the system, i.e. infinite in the thermodynamic limit, appears.
Analogously, the process can be defined on the bonds joining neighbouring sites, with the only difference that the site occupation probability $q$ is replaced by a bond occupation probability. The bond percolation corresponds to the site percolation on the conjugate lattice (obtained exchanging bonds with sites and viceversa), and gives similar results.

Recently, percolation theory has been successfully introduced in the field of complex networks, with some seminal works on undirected uncorrelated random graphs \cite{callaway,cohen2,cohen5,bollobas2}. A number of other works followed, studying
the cases of correlated \cite{newman3,newman4,vazquez,hopcroft} and directed \cite{dorogo2,schwartz,boguna,sars2} random graphs. 

The simplest example of percolation on graphs is the topological phase transition occurring in the generation of an Erd\"os-R\'enyi random graph \cite{erdos,bollobas}. Indeed, the static construction algorithm proposed by Erd\"os and R\'enyi for a random graph with $N$ vertices requires no embedding space and, for $N \rightarrow \infty$, corresponds to the mean-field bond percolation on an infinite lattice \cite{stauffer}. In this case, the bond percolation threshold is the value of edge (bond) occupation probability, at which an infinite cluster appears, namely, in the language of graph theory, a giant component emerges. The celebrated Molloy-Reed criterion for percolation \cite{molloy1,molloy2} shows that, for this model, a giant component exists if and only if the edge occupation probability is larger than $1/N$. Hence, even simple examples like this one suggest that percolation is directly related to the structural properties of graphs.

In addition, the correct functioning of real complex networks largely relies on their structural properties:  
the resilience to damage and the robustness to external attacks, cascade failures, or to collapses due to traffic jams and overloadings \cite{callaway,cohen2,cohen3,holme1,holme2,motter,watts}.
A common way to evaluate the robustness of a graph is that of studying the size $\mathcal{G}_{f}$ of the giant component after the removal of a fraction $f$ of vertices. Due to nodes removal, $\mathcal{G}_{f}$ is expected to decrease, but the remarkable point is that a random removal of nodes (also called random failure process) can be mapped exactly on a site percolation problem, with $f = 1- q$.  Percolation theory predicts the presence of a threshold value $f_c = 1- q_c$ above that the giant component disappears, i.e. the network is fragmented in many disconnected components of very small sizes ($\sim \mathcal{O}(\log N)$) with respect to the total number $N$ of vertices in the graph. 
As for lattices, also for general random networks site percolation and bond percolation (or nodes and edge percolation, using network's language) are complementary notions.  

Percolation has found applications also in the context of spreading phenomena on networks, such as epidemic models \cite{moore1,moore2,newman1,newman2}: for some of them, the connection with percolation is a longtime topic (see for example \cite{frisch,grassberger,warren,warren2}), while for other models, more rigorous descriptions, based on the analysis of population equations of epidemiology, are required \cite{may,pastor1,pastor2,boguna3,boguna2,moreno1,moreno2,pastor,boguna4}.
The models of epidemic spreading solved using percolation theory identify vertices with
individuals and edges as pairwise contacts between them \cite{newman1}. A way to distinguish disease-causing contacts from  safe ones, is that of introducing a parameter, called {\it transmissibility}, accounting for the disease's transmission probability along the edges. Without any claim of characterizing all real properties related to transmissibility, the model's behaviour is easily determined using edge percolation, that establishes the presence of a giant component as function of the transmission probability, and states the possibility for global epidemic outbreaks. 
In this context, the existence of a giant component means that we can reach an large (infinite) number of other nodes of the same graph starting by one of them and moving along the edges. 

Nervertheless, the edge percolation cannot encode all the features of a spreading process. Real spreading phenomena are more complex, they need time and resources to traverse the network and the spreading rates themselves depend on the properties of the nodes and the edges and on the details of the process. 
Many examples come from daily life: in the Internet, connections between computers are real cables, thus the transmission along them depends on their bandwidth; in air-transportation networks, the edges are weighted with the number of available seats, that is a measure of the their capacity; in social and contact networks as well, the importance of links is related to the type of the relationships between individuals.

~\\

The aim of this paper is that of developing a more general approach to spreading phenomena using percolation, endowing the nodes and the edges with some properties that quantify their ability to spread flows throughout the network. 

The idea of a percolation process in which the basic elements (nodes and edges) can present inhomogeneous properties goes back to the physical studies on the conduction in resistor networks models (see Ref.~\cite{kirkpatrick} and references therein). A typical way to introduce disorder on the edges (or on the nodes) is that of assigning to each of them a random number between $0$ and $1$, then removing all edges (nodes) in a selected interval of values. 
Percolation properties are studied as functions of the fraction of edges (nodes) that remains. Powerful methods, as that of the {\em effective medium theory} (\cite{bruggeman,landauer}) have provided a general comprehension of the large disorder regime of such processes. 
Mathematicians have studied similar types of site and bond percolation models using the theory of {\em random fields}, but  only partial or isolated results have been achieved (see Refs.~\cite{garet,yu} or Refs.~\cite{georgii,grimmett,kesten} for a general introduction). 
   
Using the well-known approach of generating functions we are able to study a class of percolation problems on random graphs that are related to the original random resistor network model and that we will call {\em inhomogeneous percolation models on random graphs}. In particular, we will study these models in relation with spreading phenomena.   

In order to express the ability of an edge $(i, j)$ to transfer some physical quantities (information, energy, diseases, etc.), we introduce an {\em edge transition probability} $T_{ij}$, generally dependent on the properties of the vertices $i$, $j$ and of the edge $(i,j)$ itself. In addition, in real processes the nodes have different ability to spread, thus they should be supplied with a {\em node traversing probability} $q_{i}$.
In this manner, the actual size of the giant component can be affected by the non-optimal or non-homogeneous flow through nodes and edges, the most general model performing an {\em  inhomogeneous joint site-bond percolation}.
On the other hand, information about simpler processes such as site percolation with inhomogeneous bonds (edges) or bond percolation with inhomogeneous sites (nodes) can be easily obtained assuming, in turn, $q_{i}$ or $T_{i j}$ as uniform.
In these particular cases we will specify them as node ($q$) or edge ($T$) occupation probability.
  
In relation with the modern theory of complex networks \cite{vespiL,dorogoL,AB,newmanR}, this inhomogeneous percolation model can also provide further information about the robustness of random and real networks.   
It allows to study the behaviour of the {\em functional giant component}, that can be defined as the part of the giant component that can bear and handle a large flow of information, ensuring the global well-functioning of the network. 
Edges with small or large transmission abilities are topologically identical, but spreading processes can be deeply influenced by the existence of fragile connections collapsing under the load of large flows.

~\\

The present paper is organized as follows. In Section~\ref{naivesec}, we propose an intuitive way of modeling a simple 
problem of random resistor networks as a site percolation with noisy edges, leading to approximated but interesting results. A rigorous approach to the general problem of inhomogeneous joint site-bond percolation, by means of the generating functions formalism, is developed in Section~\ref{mainsec}. It contains a deep analysis of the case of correlated random graphs starting from very general assumptions. The main result obtained by this technique is a generalization of the Molloy-Reed criterion for the existence of the giant component \cite{molloy1,molloy2}. 
With further assumptions, one easily recovers well-known results on the threshold value for site percolation in uncorrelated random graphs, with some interesting intermediate situations.  
Section~\ref{appsec} is devoted to applications: we derive analytically the expression of the critical threshold of site percolation for different functional forms of the transition probability, testing some of these results by numerical simulations.
Conclusions on the relevance of this approach are exposed in Section~\ref{concsec}. 
Some technical details on the derivation of formulae presented in Section~\ref{mainsec} are reported in the Appendices.

~\\

\section{INTUITIVE ARGUMENTS FOR SITE-PERCOLATION IN UNCORRELATED RANDOM GRAPHS WITH NOISY EDGES}
\label{naivesec}

In this section, we present an introductory example that can be related to the study of random resistor networks in which each edge $(i ,j)$ is given a random conductance value $T_{i j}$ between $0$ and $1$. The probability of traversing an edge is a function of this conductance. In resistor networks it is usually treated as a step function (the Heaviside $\theta(T_{i j}>T_{0})$ for a reasonable value $T_0$), so that transmission occurs only along a fraction of the edges (with sufficiently high conductance) and, for those edges, it is optimal. Here, we are interested in spreading phenomena, in which the probability of passing through an edge should rather be a linear function of $\{T_{i j}\}$; for this reason, we consider directly the set of $\{T_{i j}\}$ as the set of edge transition probabilities. A simple method allows to study this model as a site percolation with noisy edges (the name comes from mathematical research in the field of wireless communication \cite{franceschetti}) and to describe how the presence of edge transition probabilities affects general structural properties of networks. 

We use an intuitive derivation of the percolation criterion that follows directly from the approach introduced by Cohen et al. in Ref.~\cite{cohen2} for uncorrelated sparse random graphs. 
Let us call ${\rho}_{ij}$ the probability that an edge in the network does not lead to a vertex connected via the remaining edges to the giant component (infinite cluster) and $T_{ij}$ the probability that a flow leaving the node $i$ passes trough the edge $(i,j)$, reaching the node $j$. Both ${\rho}_{ij}$ and $T_{ij}$ are defined in the interval $[0,1]$.

Note that the definition of transition probabilities resembles the one originally introduced by Newman in Ref.~\cite{newman1,newman2} with the name of transmission probability (or transmissibility) to study epidemic spreading. For this reason, in the following, we will sometimes refer to the transition probability as the transmissibility of an edge.    

\begin{figure} 
\centerline{
\begin{tabular}{|c|}\hline \\ \includegraphics*[angle=-90,width=0.45\textwidth]{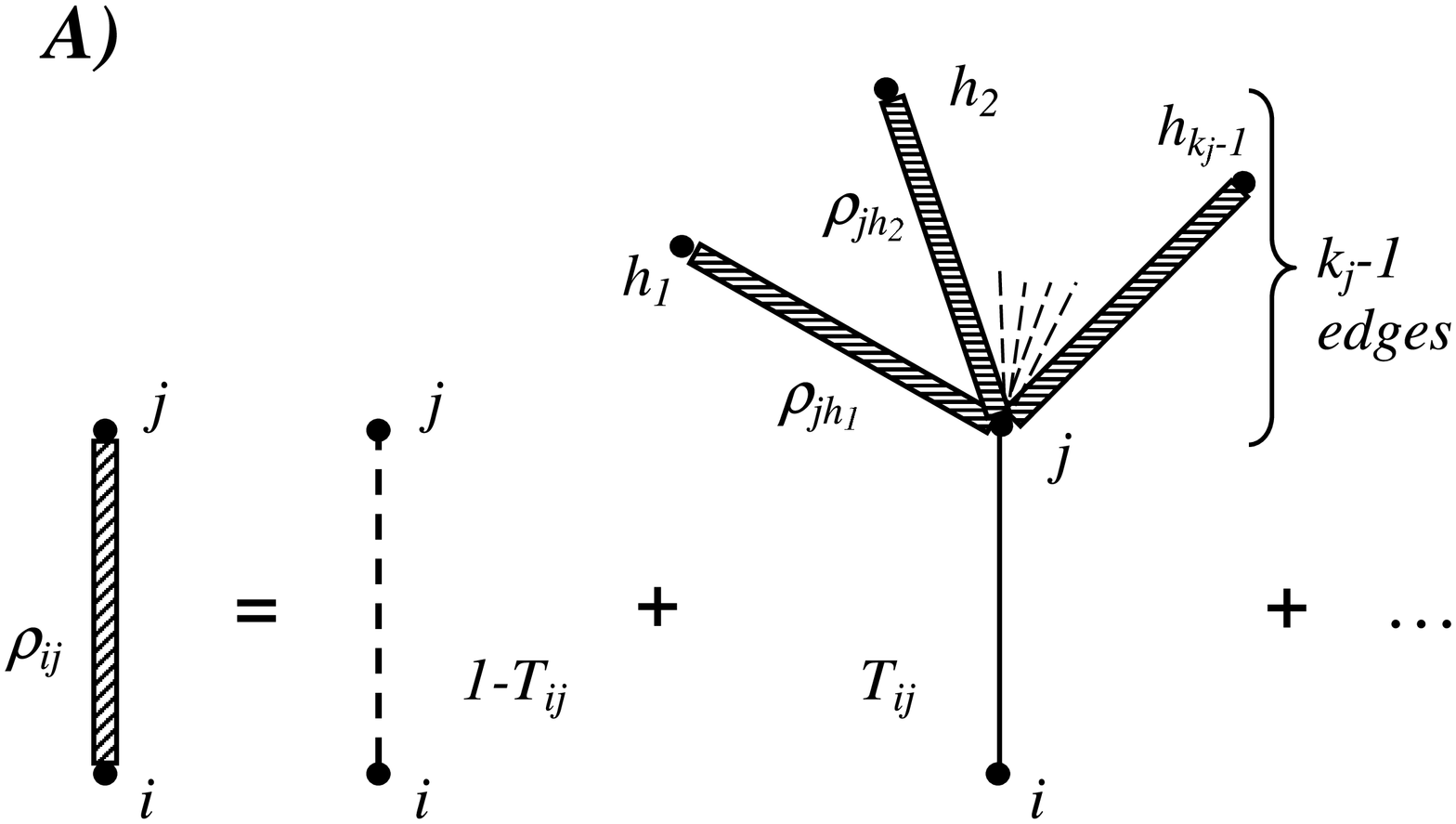}\\ \hline\end{tabular}~~
\begin{tabular}{|c|}\hline \\ \includegraphics*[angle=-90,width=0.45\textwidth]{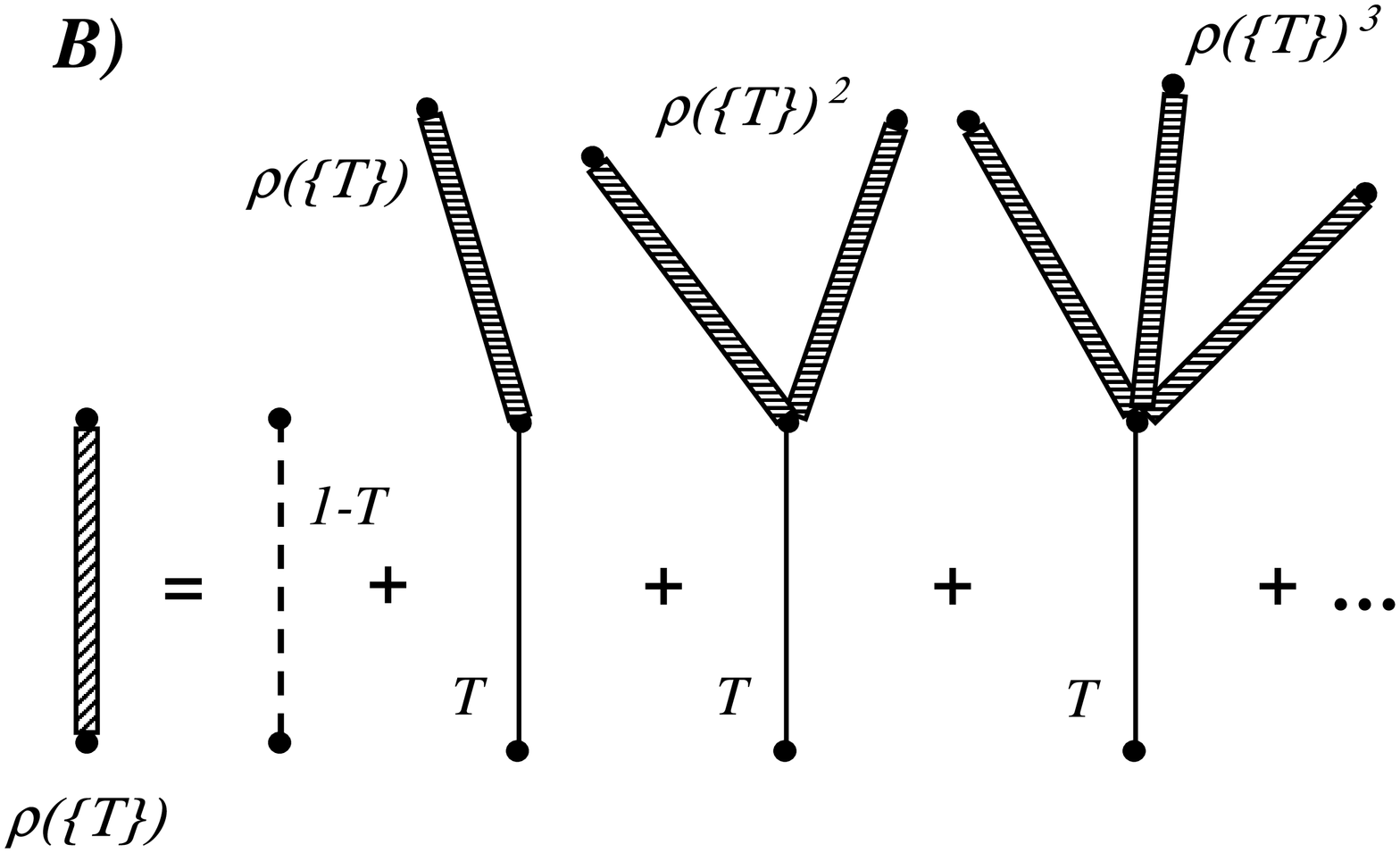}\\ \hline\end{tabular}
}

\caption{A) Graphical representation of the Eq.~\ref{naive0}. Fat striped lines indicate the contribution coming from the unknown probability $\rho_{ij}$ that an edge $(i, j)$ does not belong to the infinite cluster. The first term on the right-hand side represents the contribution coming from the probability $1-T_{ij}$ that a flow does not traverse the edge $(i, j)$; the second term accounts for the probability that the edge $(i, j)$ is traversed times the unknown contribution $\rho_{jh}$ of other edges attached to $j$.\\
B) Graphical representation of the self-consistent Eq.~\ref{naive1}. Fat striped lines correspond to the unknown uniform probability $\rho(\{T\})$ that an edge does not lead to the giant component. Full lines mean that edges can be crossed, with probability $T$, while the dashed edge corresponds to the contribution $1-T$ that an edge cannot be crossed.    
}
\label{naive_ex}
\end{figure}

In general, ${\rho}_{ij}$ depends on the transition probability $T_{ij}$ of the edge $(i,j)$, by a factor $1- T_{ij}$, and on the probability that the node $j$, if reached, does not belong to the infinite cluster.
This contribution can be computed as the joint probability that each one of the remaining edges emerging from $j$ does not belong to the giant component.  
The different contributions are assembled as in Fig.~\ref{naive_ex}-A, giving the recursive expression
\begin{equation}
{\rho}_{ij} = 1-T_{ij} + T_{ij} \prod_{{\scriptsize \begin{array}{c} h\in V(j)\\ h\neq i \end{array}}} {\rho}_{jh}~,
\label{naive0}
\end{equation}
that contains a product of $k_{j}-1$ terms $\rho_{jh}$ if $k_{j}$ is the degree of the node $j$ and $V(j)$ is its set of  neighbours.
This recursive procedure is not closed, but the introduction of some hypotheses on the set of transition probabilities $\{T\}$ allows a statistical reformulation of Eq.~\ref{naive0}.
Let us suppose that the transition probabilities are random variables with a given ditribution $p_{T}$, and the average overall probability $\rho$ that a randomly chosen edge does not belong to the giant component depends only on some general properties of that distribution such as the mean, the variance, etc. (i.e. $\rho = \rho(\{T\})$).   
Moreover, by picking up an edge at random in a random graph, the probability that it is connected to an edge of degree $k$ is $\sum_{k} \frac{k p_{k}}{\langle k \rangle}$, where $p_{k}$ is the degree distribution of the graph. 
We can now write a self-consistent equation for $\rho (\{T\})$,
\begin{equation}
\rho (\{T\}) = 1-T + T \sum_{k}\frac{k p_{k}}{\langle k \rangle} {\rho (\{T\})}^{k-1}~,
\label{naive1}
\end{equation}
in which $T$ is a realization of the independent and identically distributed (i.i.d.) random variables $\{T\}$ in the interval $[0,1]$. Averaging both the members of the Eq.~\ref{naive1} on $p_{T}$, we obtain that the probability $\rho$ depends only on the average value $\langle T \rangle$, i.e. $\rho (\{T\})=\rho (\langle T\rangle)$, and the equation becomes
\begin{equation}
\rho (\langle T\rangle) = 1- \langle T\rangle + \langle T \rangle \sum_{k} \frac{k p_{k}}{\langle k \rangle} {\rho (\langle T\rangle)}^{k-1} = I[\rho (\langle T\rangle)]~.
\label{eq_1b}
\end{equation}
The equation is graphically represented in Fig.~\ref{naive_ex}-B.  
Apart from the trivial solution $\rho =1$, another solution $\rho ={\rho}^{*}<1$ exists if and only if $\frac{d I}{d \rho}{\vert}_{\rho =1} \geq 1$. The curve $I[\rho]$, indeed, is positive in $\rho=0$ ($I[0] = 1-\langle T \rangle + \langle T\rangle p_{1} /\langle k\rangle > 0$), therefore $\frac{d I}{d \rho}{\vert}_{\rho =1} \geq 1$ means that it crosses the line $f(\rho)=\rho$ in a point $0 < {\rho}^{*} < 1$.
The condition on the derivative of the r.h.s. of Eq.~\ref{eq_1b} corresponds to
\begin{equation} 
\langle T \rangle \frac{\langle k^{2} \rangle - \langle k \rangle}{\langle k \rangle} \geq 1~.
\label{condgc}
\end{equation}
A generalization of the Molloy-Reed criterion \cite{molloy1} for the existence of a giant component in presence of noisy edges immediately follows,
\begin{equation}
\frac{\langle k^2 \rangle}{\langle k \rangle} \geq 1+\frac{1}{\langle T \rangle}~,
\label{mrgc}
\end{equation}
meaning that, when the transition probabilities are i.i.d. random variables, a giant component exists if and only if the inequality is satisfied. The case of uniform transition probabilities is exactly the same, with uniform value $T$ instead of the average value $\langle T \rangle$.
Note that, while in the case of perfect transmission ($T_{ij}=1$) the usual formulation of the Molloy-Reed criterion is recovered \cite{molloy1}, when $\langle T \rangle < 1$ the r.h.s. of Eq.~\ref{mrgc} can grow considerably, affecting the possibility of observing percolation: the smaller is the average transition probability, and the larger are connectivity fluctuations $\langle k^2 \rangle$ needed to ensure the presence of a giant component.

\begin{figure} 
\centerline{
\includegraphics*[angle=-90,width=0.5\textwidth]{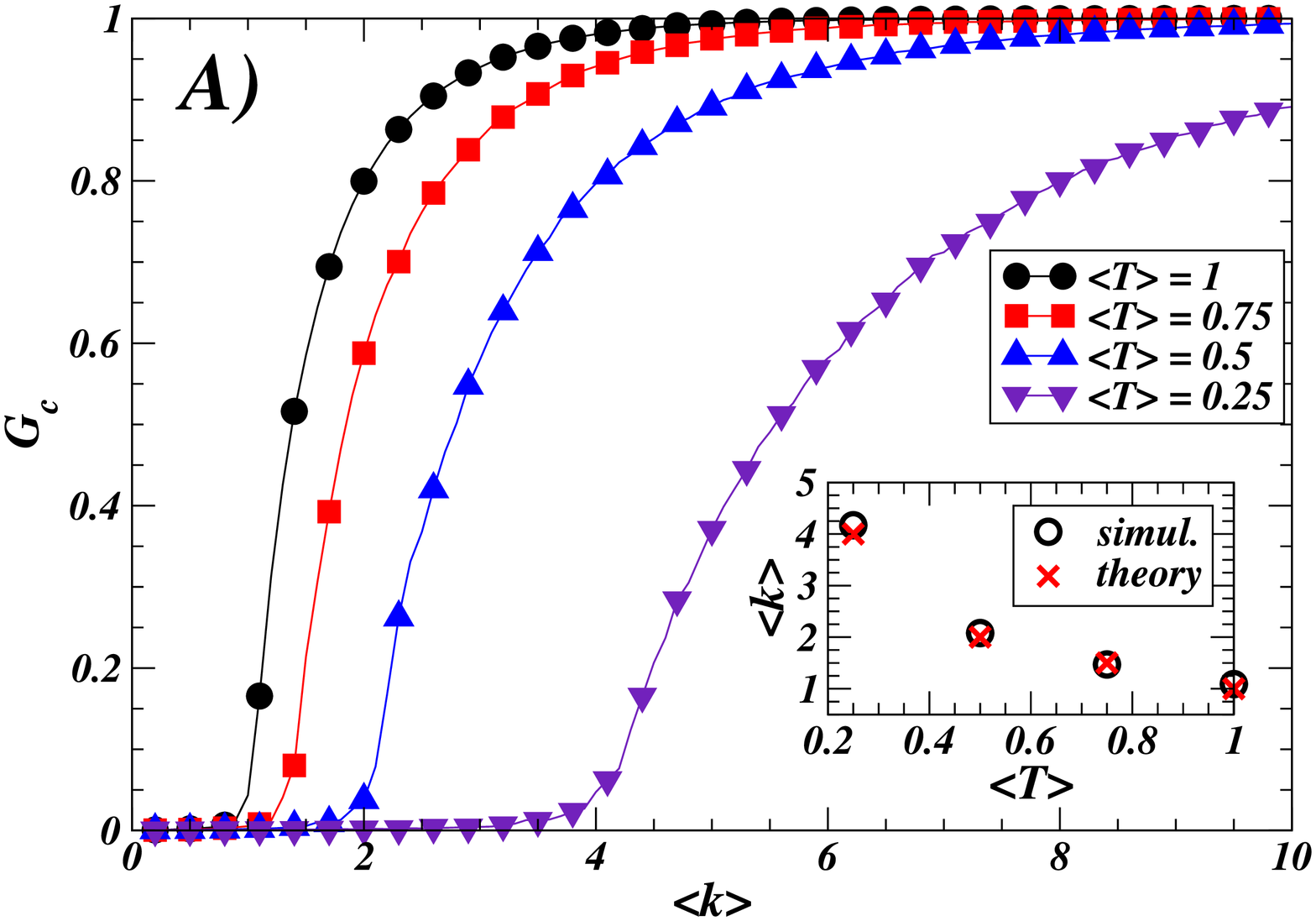}~~
\includegraphics*[angle=-90,width=0.5\textwidth]{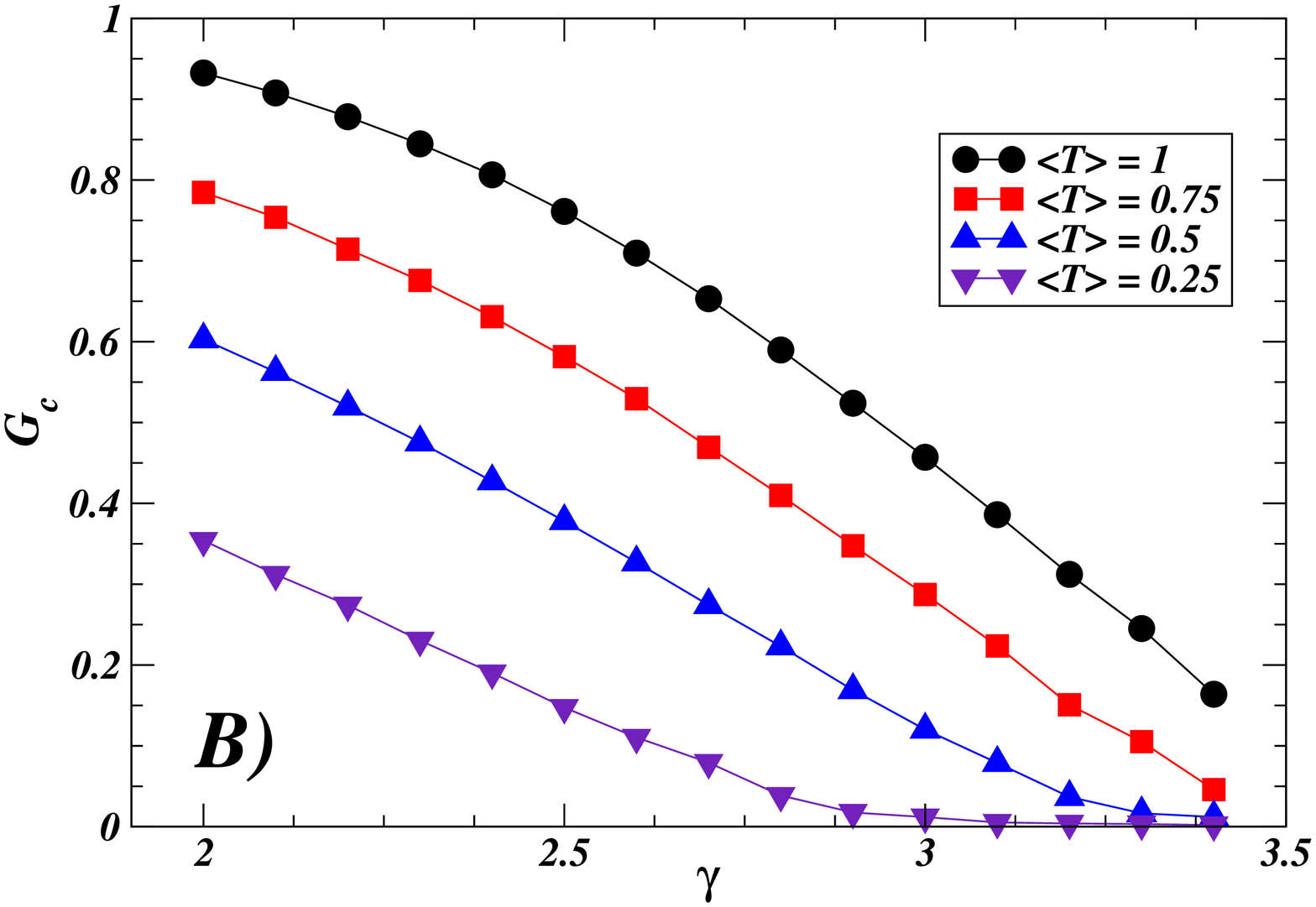}
}
\caption{A) Size of the giant component $\mathcal{G}_{c}$ vs. average degree $\langle k\rangle$ in an  Erd\"os-R\'enyi random graph in which the edge transition probabilities are uniformly random distributed with mean value $\langle T\rangle$. Simulations are performed on a sample of $100$ graphs with $N=10^4$ nodes. The different curves refer to different values of average transition probability: $\langle T \rangle = 1$ (circles), $0.75$ (squares), $0.5$ (up triangles) and $0.25$ (down triangles). The inset reports the numerical values (circles) of $\langle k \rangle$ at which the giant component appears for different $\langle T \rangle$ along with the values predicted by the theory (crosses).\\ 
B) Size of the giant component $\mathcal{G}_{c}$ vs. the exponent $\gamma$ in an power-law graph in which the edge transition probabilities are uniformly random distributed with mean value $\langle T\rangle$.  Simulations are performed on a sample of $100$ graphs with $N=10^4$ nodes, cut-off $\kappa = 10^2$ and minimum degree $m=1$. The different curves refer to different values of average transition probability: $\langle T \rangle = 1$ (circles), $0.75$ (squares), $0.5$ (up triangles) and $0.25$ (down triangles).\\}
\label{bound}
\end{figure}

For a random graph with poissonian degree distribution (i.e. $p_{k} = e^{-\langle k \rangle} {\langle k \rangle}^k/k!$), the criterion in Eq.~\ref{mrgc} corresponds exactly to have $\langle k \rangle \geq 1/ \langle T \rangle$.
We show in Fig.~\ref{bound}-A the results of numerical simulations of the giant component's computation on an Erd\"os-R\'enyi random graph with $N = 10^4$ nodes and with random transition probabilities uniformly distributed in $[a,b]$ with $0 \leq a < b \leq 1$. A giant component clearly appears when the mean connectivity $\langle k \rangle$ exceeds the inverse of mean transmissibility value $(b-a)/2$. Considering different distributions (e.g. binomial distributions) for the i.i.d. random variable $T$ does not affect the results.

%
%
%
%
For heterogeneous graphs with broad degree distribution, the inequality in Eq.~\ref{mrgc} is satisfied thanks to the huge fluctuations of the node degrees ensuring the l.h.s. to be larger than $1+ 1/ \langle T \rangle$. In particular, when the graph has power-law degree distribution $p_{k} \sim m k^{-\gamma}$ ($2 < \gamma < 3$, $m$ is the minimum degree), fluctuations diverge in the limit $N \rightarrow \infty$. In this case, the giant component always exists if $N$ is large enough. If $\gamma > 3$, the second moment is finite, and Eq.~\ref{mrgc} provides the following bound for the average transition probability necessary to have a giant component (computed using the continuum approximation for the degree),
\begin{equation}
{\langle T \rangle} \geq \frac{\gamma-3}{\gamma (m-1) + 3 - 2m}~.
\label{sfth}
\end{equation}
The inequality is always satisfied for $m \geq 1+1/ \langle T\rangle$, while for $m < 1 + 1/ \langle T \rangle$ it is satisfied when $\gamma \leq 3+ \frac{m\langle T\rangle}{1-(m-1)\langle T \rangle}$. For $m=1$, an infinite (uncorrelated) scale-free graph presents a giant component of order $\mathcal{O}(N)$ only when $\gamma \leq 3 + \langle T \rangle \leq 4$. 
At $\gamma = 3$, logarithmic corrections should be taken into account \cite{cohen2}.

Since real networks are large but finite, it is important to compute the finite-size effects for power-law graphs with exponent $2 < \gamma < 3$, introducing a cut-off $\kappa(N)$ on the degree. Following Ref.~\cite{cohen2}, the expression  for $\langle k^{2} \rangle / \langle k \rangle$ can be easily computed in the continuum approximation for the degree, obtaining an explicit relation for the bound on $\gamma$ and $\langle T \rangle$,
\begin{equation}
\langle T \rangle \geq \frac{1}{(\frac{\gamma-2}{3-\gamma}) m {\kappa(N)}^{3-\gamma}}~.       
\end{equation}
The r.h.s. has a maximum at $\gamma = 2+ 1/\log(\kappa)$, in which it assumes the value  
\begin{equation}
\frac{\log(\kappa)-1}{m {\kappa}^{\frac{\log(\kappa)-1}{\log(\kappa)}} + 1-\log(\kappa)}~.
\end{equation}
For $m=1$, this expression reaches a maximum value $\simeq 0.6$ for $\kappa \simeq 10$, then it decreases monotonously for larger values of the cut-off. For $\kappa \simeq 10^2$, it reduces to $0.1$, showing that the existence of a giant component can be ensured even for small $\langle T\rangle$ if the cut-off is sufficiently large.
However, in Fig.~\ref{bound}-B we have reported the size of the giant component for power-law random graphs as a function of the exponent $\gamma$ for different values of the average edge transition probability $\langle T \rangle$. The simulations are performed on graphs of $N =10^4$ nodes with cut-off $\kappa=10^2$.  They clearly show that in finite power-law networks, the real size of the giant component is dramatically affected by low transition probabilities.

%
%
%
%
Note that inverting this generalization of Molloy-Reed criterion (Eq.~\ref{mrgc}), and looking at the critical value of average transition probability ${\langle T \rangle}_{c}$ that is sufficient for the graph to admit a giant component (as we have done in Eq.~\ref{sfth}) gives results completely equivalent to those obtained by standard (homogeneous) bond percolation with $\langle T \rangle$ equal to the uniform edge occupation probability $T$ (see, for example, Ref.~\cite{newman1}).

Finally, the argument of Cohen et al. \cite{cohen2} can be used to compute the threshold of site percolation with noisy edges. If $q$ is the (uniform) node occupation probability in an uncorrelated random graph, then a randomly chosen node, whose natural degree in case of $q=1$ is $k'$, will assume a degree $k \leq k'$ with probability 
\begin{equation}
\left(\begin{array}{c} k' \\ k \end{array} \right) {q}^{k} {(1-q)}^{k'-k}~,
\end{equation}
that means that the corresponding degree distribution $P_{q}(k)$ is related to the degree distribution $P_{q=1}(k')$ by 
\begin{equation}
P_{q}(k) = \sum_{k'=k}^{\infty} P_{q=1}(k') \left(\begin{array}{c} k' \\ k \end{array} \right) {q}^{k} {(1-q)}^{k'-k}~.
\end{equation}
It follows that ${\langle k \rangle}_{q} = q {\langle k \rangle}_{q=1}$ and ${\langle k^2 \rangle}_{q} = q {\langle k^2 \rangle}_{q=1} +q(1-q){\langle k \rangle}_{q=1}$, that introduced into the expression Eq.~\ref{mrgc} for the Molloy-Reed criterion gives the expression of the threshold value for site percolation with noisy edges,
\begin{equation}
q_{c} = \frac{1}{\langle T \rangle}\frac{\langle k \rangle}{\langle k^2 \rangle - \langle k \rangle}~.     
\label{thres0}
\end{equation}
Eq.~\ref{thres0} shows that decreasing the average transmission capacity of the edges (i.e. edge transition probabilities) causes a rise of the value of node occupation probability necessary to ensure percolation.

The intuitive method we have proposed in this section is correct for uncorrelated graphs where the edges are supplied with uniform or homogeneously randomly distributed transition probabilites. The next section is devoted to the formalism of generating functions that is appropriate for deriving more general results for the case of two-point correlated random graphs.  

~\\

\section{GENERAL THEORY OF INHOMOGENEOUS PERCOLATION IN CORRELATED RANDOM GRAPHS}
\label{mainsec}

Percolation theory on graphs has been formally developed by means of generating functions \cite{callaway,dorogoR,wilf}, a method that can be successfully applied also in the present case of inhomogeneous joint site-bond percolation. Since in general we will consider edge transition probabilities depending on the properties of the two extremities of the edges, the correlations between neighbouring nodes should be taken into account. 
More precisely, a very general type of generating functions is introduced in order to describe the properties of random graphs with vertex-vertex pair correlations (also called Markovian networks \cite{boguna2}) and with heterogeneously distributed transition probabilities on the edges. 
Our main result is a general version of the Molloy-Reed criterion for the existence of a giant component in the case of joint site-bond percolation and, consequently, the expressions of the critical threshold for the two separate cases of site percolation and bond percolation.
We conclude the section discussing some examples in which the functional form of the transition probabilities allows  explicit calculations of the percolation threshold. Our notation is the same as in recent expositions (see Refs.~\cite{callaway,dorogoR}), nevertheless we refer to Appendix-\ref{append1} for a brief introduction to generating functions in random graphs.

\subsection{Generating functions approach for Markovian networks}

In graph theory, two-points correlations are usually expressed as functions of the degree by the {\em degree conditional probability} $p(k'|k)$ that is the probability that a vertex of degree $k$ is connected to a vertex of degree $k'$. This has led to the definition of a class of correlated networks, called undirected {\em Markovian} random networks \cite{boguna2} and defined only by their degree distribution $p_{k}$ and by the degree conditional probability $p(k'|k)$. 
The function $p(k' |k)$ satisfies the normalization constraint 
\begin{equation}
\sum_{k'} p(k'|k) = 1~,
\label{pr1}
\end{equation} 
and a detailed balance condition \cite{boguna2}
\begin{equation}
k p(k'|k) p_{k} = k' p(k| k') p_{k'}~. 
\label{pr2}
\end{equation}
In uncorrelated graphs, $p(k' |k)$ does not depend on $k$ and it can be easily obtained inserting Eq.~\ref{pr1} into Eq.~\ref{pr2}, 
\begin{equation}
p(k' | k) = \frac{k' p_{k'}}{\langle k \rangle}~.
\label{fact}
\end{equation}

In this context, we can assume that also the edge transition probability $T_{ij}$ depends only on the degrees $k_{i}$ and $k_{j}$ of the extremities.
Note that, while the analysis of the standard site (bond) percolation is based on the relation between the degree distribution $p_{k}$ and the degree-dependent node occupation probability $q_{k}$, in the inhomogeneous joint site-bond percolation the relation is between the pair of distributions $\{p_{k}, p(k'|k)\}$ and the pair of probability functions $\{q_{k}, T_{k k'}\}$.

A simple approach to inhomogeneous percolation is that of considering an approximated expression for the probability that an edge emerging from a vertex of degree $k_{i}$ is traversed by the flow. By averaging over the contributions of all the second extremities of the edges emerging from a node, we obtain a degree-dependent {\em effective average transmission coefficient}  $\tau_{k_{i}} = \sum_{k_{j}} T_{k_{i} k_{j}} p(k_{j}|k_{i})$ ($0 \leq \tau_{k_{i}} \leq 1$). 
The advantage of this method is that of providing analytically solvable equations for the percolation condition. Moreover, even if the sum on $k_{j}$ introduces an approximation, the effective term $\tau_{k_{i}}$  does not neglect the contributions due to the edge transition probabilities, weighting them with the correct degree conditional probability as required for Markovian networks. If these contributions are similar the approximation is very good, while when the edge transition probabilities or the nodes correlations are highly heterogeneous a different analysis should be applied (see Section~\ref{multistate}). 

According to this approximation, the generating function $F_{0}(x; \{q, T\})$ of the probability that a spreading process emerging from an occupied node flows through exactly $m$ nodes (whatever their degrees are) is written as  
\begin{equation}
F_{0}(x; \{q, T\}) = \sum_{k_{i}=1}^{\infty} p_{k_{i}} q_{k_{i}} {\left[1+ (x-1) \sum_{k_{j}=1}^{\infty} T_{k_{i} k_{j}} p(k_{j}|k_{i}) \right]}^{k_{i}}~. 
\label{ff0}
\end{equation}
This expression can be derived using simple arguments. The probability that exactly $m$ of the $k_{i}$ edges rooted in $i$ are connected to vertices of degree $k_{j}$ and that they are {\em open} (i.e. they let the flow pass along them) is 
\begin{equation}
\left(\begin{array}{c} k_{i} \\ m \end{array}\right) [T_{k_{i} k_{j}} p(k_{j}|k_{i})]^{m} {[1 - T_{k_{i} k_{j}}p(k_{j}|k_{i})]}^{k_{i}-m}~. 
\label{probcond0}
\end{equation}
If we average over the contributions of all possible second extremities, $T_{k_{i} k_{j}} p(k_{j}|k_{i})$ is replaced by $\sum_{k_{j}} T_{k_{i} k_{j}}p(k_{j}|k_{i})$; then putting together these terms with an $x$ variable for each open edge, summing over all the possible degree $k_{i}$ and over all the possible values of $m$, we get 
\begin{equation}
\begin{split}\label{ff0bis}
F_{0}(x; \{q, T\})& = \sum_{m=0}^{\infty} \sum_{k_{i}=m+1}^{\infty} p_{k_{i}} q_{k_{i}} \left(\begin{array}{c} k_{i} \\ m \end{array}\right) {\left[\sum_{k_{j}} T_{k_{i} k_{j}} p(k_{j}|k_{i})\right]}^{m} {\left[1 - \sum_{k_{j}} T_{k_{i} k_{j}}p(k_{j}|k_{i})\right]}^{k_{i}-m} x^{m} \\
&= \sum_{k_{i}=1}^{\infty} \sum_{m=0}^{k_{i}} p_{k_{i}} q_{k_{i}} \left(\begin{array}{c} k_{i} \\ m \end{array}\right) {\left[x \sum_{k_{j}} T_{k_{i} k_{j}} p(k_{j}|k_{i})\right]}^{m} {\left[1 - \sum_{k_{j}} T_{k_{i} k_{j}}p(k_{j}|k_{i})\right]}^{k_{i}-m} \\
&= \sum_{k_{i}=1}^{\infty} p_{k_{i}} q_{k_{i}} {\left[1+ (x-1) \sum_{k_{j}=1}^{\infty} T_{k_{i} k_{j}} p(k_{j}|k_{i}) \right]}^{k_{i}}~. 
\end{split}
\end{equation} 
The value $F_{0}(1;\{q,T\})$ assumed by this function in $x=1$ is the average occupation probability $\langle q \rangle = \sum_{k} p_{k} q_{k}$, that is consistent with the fact that summing over the contributions of all the possible amounts of emerging edges traversed by the flow means that we are simply considering the number of starting nodes, i.e. the average number of occupied nodes. 
The first derivative with respect to $x$ computed in $x=1$
\begin{equation}
\frac{\partial}{\partial x}F_{0}(x; \{q, T\}){\bigg \vert}_{x=1} = \sum_{k_{i}=1}^{\infty} k_{i} p_{k_{i}} q_{k_{i}} \sum_{k_{j}=1}^{\infty} T_{k_{i} k_{j}} p(k_{j}| k_{i})~,
\end{equation}
is the average number of open edges emerging from an occupied vertex.

Using similar arguments, we get the generating function $F_{1}(x; \{q, T\})$ of the probability that the flow spreading from a vertex, reached as an extremity of an edge picked up at random, passes through a given number of the remaining edges,
\begin{equation}
F_{1}(x; \{q, T\}) = \sum_{k_{i}=1}^{\infty} \frac{k_{i}  p_{k_{i}}}{\sum_{k} k p_{k}} q_{k_{i}} {\left[1 + (x-1)\sum_{k_{j}=1}^{\infty} T_{k_{i} k_{j}} p(k_{j}| k_{i}) \right]}^{k_{i}-1}~. 
\label{ff1}
\end{equation}

Following the notation presented in Appendix-\ref{append1}, we call $H_{0}(x; \{q, T\})$ the generating function of the probability that a randomly chosen vertex belongs to a cluster of given size and $H_{1}(x; \{q, T\})$ the generating function of the probability that a randomly chosen edge leads to a vertex belonging to a cluster of a given size. They satisfy a system of self-consistent equations 
\begin{subequations}
\begin{align}
H_{0}(x; \{q, T\})& = 1 - F_{0}(1; \{q, T\}) + x F_{0}\left[ H_{1}(x; \{q, T\}); \{q, T\} \right]~, 
\label{apph0}\\
\nonumber ~\\
H_{1}(x; \{q, T\})& = 1 - F_{1}(1; \{q, T\}) + x F_{1}\left[H_{1}(x; \{q, T\}); \{q, T\}\right]~. 
\label{apph1}
\end{align}
\end{subequations} 

The expression for the mean cluster size $\langle s \rangle$ is obtained by derivation of Eq.~\ref{apph0}, 
\begin{equation}
\langle s \rangle = \frac{d}{d x} H_{0} (x; \{q, T\}) {\bigg \vert}_{\tiny x=1} = H_{0}' (1; \{q, T\})~,  
\end{equation}
where $x=1$ implies the average over all possible degrees $k_{i}$.
From Eq.~\ref{apph0}-\ref{apph1}, using the relation $H_{1}(1; \{q, T\}) =1$, we get 
\begin{equation}
\langle s \rangle = F_{0}(1; \{q, T\}) + F_{0}'(1; \{q, T\}) H_{1}'(1; \{q, T\})~,
\label{apps1}
\end{equation}
where the expression for $H_{1}'(1; \{q, T\})$ comes directly from deriving Eq.~\ref{apph1} with respect to $x$ in $x=1$,
\begin{equation}
H_{1}'(1; \{q, T\}) = \frac{F_{1}(1; \{q, T\})}{1 - F_{1}'(1; \{q, T\})}~.
\end{equation}
Inserted into Eq.~\ref{apps1} it leads to the well-known expression (\cite{callaway})
\begin{equation}
\langle s \rangle = F_{0}(1; \{q, T\}) + \frac{F_{0}'(1; \{q, T\})}{1 - F_{1}'(1; \{q, T\})}~;
\label{apps2}
\end{equation}
hence, the giant component exists if and only if $F_{1}'(1; \{q, T\}) \geq 1$.
Explicitly, it means that the generalized Molloy-Reed criterium for the inhomogeneous joint site-bond percolation on Markovian random graphs takes the following form, 
\begin{equation}
\sum_{k_{i}, k_{j} = 1}^{\infty} p_{k_{i}} k_{i} \left[(k_{i} -1) q_{k_{i}} T_{k_{i} k_{j}} - 1 \right] p(k_{j}| k_{i}) \geq 0~.
\label{mrcgg}
\end{equation}
The threshold for the site percolation with inhomogeneous edges occurs when the equality holds. Inserting uniform occupation probability $q_{k} \equiv q$, its critical value is  
\begin{equation}
q_{c} = \frac{\langle k \rangle}{\sum_{k_{i}=1}^{\infty} k_{i} (k_{i}-1) p_{k_{i}} \sum_{k_{j} = 1}^{\infty} T_{k_{i} k_{j}} p(k_{j}| k_{i})}~.
\label{thresh1}
\end{equation}

However, in many natural, social and technological networks, the nodes represent individuals or organizations with very heterogeneous properties that are completely neglected looking just at a degree-level. For instance, in the Internet there
are some non-physical relationships between the nodes (Autonomous Systems) due to policies or administrative strategies that have a deep influence on the spreading of information. Other examples come from social networks, where rather homogeneous intra-community environments contrast with very heterogeneous properties exhibited switching to other groups of individuals. 
These extra-factors cannot be easily incorporated into a statistical analysis and represent a challenge for the present research in complex networks. 

A natural attempt to include these properties in the formalism of degree-dependent generating functions consists of admitting the presence of different types of nodes, as assumed by Newman in Ref.~\cite{newman4}. In the next section we derive some general results in the case of networks with {\em multi-state nodes}. 

\subsection{Generalization to networks with multi-state nodes}\label{multistate}

Suppose that the nodes of a graph are divided into $n$ classes, each one specified by a particular quality or state. Then, we consider the degree distribution $p_{k}^{(h)}$ of nodes of class $h=1, \dots, n$, conventionally normalized on the relative set of nodes, i.e. $\sum_{k} p_{k}^{(h)} = 1$. This condition ensures the normalization to $1$ for the generating functions.   
Inside the classes there are no restrictions on the transition probabilities and they might be very different. 

Without entering the details of the derivation, that are reported in Appendix-\ref{append2}, we consider the general case of {\em inhomogeneous joint site-bond percolation in multi-state correlated random graphs}.  
The generating function of the probability that a physical quantity, spreading from a vertex of class $h$, flows
through a certain number of edges and reaches vertices in the same or in different classes is 
\begin{equation}
F_{0}^{(h)}(x_{1},x_{2}, \dots, x_{n};\{q, T\}) = F_{0}^{(h)}(\mathbf{x};\{q, T\}) = \sum_{k_{i}=1}^{\infty} p_{k_{i}}^{(h)} q_{k_{i}}^{(h)} {\left[ 1+ \sum_{l=1}^{n} (x_{l}-1) \sum_{k_{j_{l}}} T^{(h\rightarrow l)}_{k_{i} k_{j_{l}}} p^{(h\rightarrow l)}(k_{j_{l}}|k_{i}) \right]}^{k_{i}}, 
\label{pre_multi3}
\end{equation}
where $q_{k_{i}}^{(h)}$ is the occupation probability of a node of class $h$ and degree $k_{i}$, $T^{(h\rightarrow l)}_{k_{i} k_{j_{l}}}$ is the transition probability from a node of class $h$ and degree $k_{i}$ to a node belonging to the class $l$ and having degree $k_{j_{l}}$, and $p^{(h\rightarrow l)}(k_{j_{l}}|k_{i})$ is the degree conditional probability in multi-state markovian graphs.
Similarly we compute the generating function $F_{1}^{(h)}(\mathbf{x};\{q,T\})$ of the probability that a randomly chosen edge leads to a vertex of degree $k_{i}$, belonging to the class $h$, and from which the spread flows towards a given number of neighbours (of the same or different classes). It reads 
\begin{equation}
F_{1}^{(h)}(x_{1},x_{2}, \dots, x_{n};\{q, T\}) = F_{1}^{(h)}(\mathbf{x};\{q, T\}) = \sum_{k_{i}=1}^{\infty} \frac{k_{i} p_{k_{i}}^{(h)}}{\sum_{k} k p_{k}^{(h)}} q_{k_{i}}^{(h)} {\left[ 1+ \sum_{l=1}^{n} (x_{l}-1) \sum_{k_{j_{l}}} T^{(h\rightarrow l)}_{k_{i} k_{j_{l}}} p^{(h\rightarrow l)}(k_{j_{l}}|k_{i}) \right]}^{k_{i}-1}. 
\label{pre_multi4}
\end{equation}
The two above quantities allow to write a system of self-consistent equations for the generating function $H_{0}^{(h)}(x;\{q, T\})$ and $H_{1}^{(h)}(x;\{q, T\})$ of the probability that a randomly chosen vertex of class $h$ or an edge leading to that node belong to a connected component of a given size, 
\begin{subequations}
\begin{align}
H_{1}^{(h)}(x; \{q, T\})& = 1-F_{1}^{(h)}(\mathbf{1};\{q, T\}) + x~ F_{1}^{(h)}[H_{1}^{(1)}(x;\{q, T\}), \dots, H_{1}^{(n)}(x;\{q, T\});\{q, T\}]~,
\label{pre_multi5}\\
\nonumber ~\\
H_{0}^{(h)}(x; \{q, T\})& = 1-F_{0}^{(h)}(\mathbf{1}; \{q, T\}) + x~ F_{0}^{(h)}[H_{1}^{(1)}(x; \{q, T\}), \dots, H_{1}^{(n)}(x; \{q,T\}); \{q, T\}]~.
\label{pre_multi6}
\end{align}
\end{subequations}
The condition determining the existence of a giant component is again the divergence of the the mean cluster size, and corresponds to the following {\em generalized Molloy-Reed criterion} for inhomogeneous joint site-bond percolation in multi-state Markovian correlated random graphs (see Appendix-\ref{append2} for details): 
\begin{equation}
\det \left[I - \mathcal{F} \right] \leq 0~,
\label{pre_multi13}
\end{equation}
where $\mathcal{F} = \nabla_{x} \mathbf{F}_{1}[\mathbf{x}=\mathbf{1}; \{q, T\}]$ (whose elements are of the type 
$\frac{\partial}{\partial x_{l}} F_{1}^{(h)}[1;\{q, T\}]$).

It is evident that such a general result strongly depends on which kind of node partition we are considering.
Two examples of partitions are particularly relevant: a single class collecting all the nodes and a degree-based classification of the nodes.

In the latter situation, each class gathers all vertices with a given degree and there is in principle an infinite number $n$ of states (in finite networks there are as many states as the number of different degrees). Since  in this case $p_{k_{i}}^{(k_{i})}=1$ and $\frac{k_{i} p_{k_{i}}^{(k_{i})}}{\sum_{k} k p_{k}^{(k_{i})}} =1$, Eqs.~\ref{pre_multi3}-\ref{pre_multi4} become 
\begin{subequations}\label{multi145}
\begin{align}
F_{0}^{(k_{i})}(x_{1}, \dots, x_{n}; \{q, T\})& = q_{k_{i}} {\left[ 1+ \sum_{k_{j}} (x_{k_{j}} -1) T_{k_{i} k_{j}} p(k_{j}|k_{i})\right]}^{k_{i}},\label{multi14}\\
F_{1}^{(k_{i})}(x_{1}, \dots, x_{n}; \{q, T\})& = q_{k_{i}} {\left[ 1+ \sum_{k_{j}} (x_{k_{j}} -1) T_{k_{i} k_{j}} p(k_{j}|k_{i})\right]}^{k_{i}-1}.
\label{multi15}
\end{align}
\end{subequations}

The self-consistent system of equations for the generating functions of clusters size probability looks like in Eqs.~\ref{pre_multi5}-\ref{pre_multi6}, and the condition for the existence of a giant component is still the divergence of the mean cluster size, but now the elements of the matrix $\mathcal{F}$ are
\begin{equation}
\mathcal{F}_{ij} = {(\nabla_{x} \mathbf{F}_{1} [\mathbf{x=1}; \{q, T\}])}_{ij} = (k_{i}-1) q_{k_{i}} T_{k_{i} k_{j}} p(k_{j}|k_{i})~.
\label{multi16}
\end{equation}
The generalized Molloy-Reed criterion becomes
\begin{equation}
\det \left[ (k_{i}-1) q_{k_{i}} T_{k_{i} k_{j}} p(k_{j}|k_{i}) -\delta_{ij} \right] \geq 0~.
\label{multi17}
\end{equation}
This expression corresponds to the criterion of Ref.~\cite{vazquez} for the existence of percolation in correlated random graphs (apart from a matrix transposition), but with the difference that in this case we are dealing with inhomogeneous joint site-bond percolation, then both the degree-dependent node traversing probability and edge transition probability appear in the expression. The condition that percolation threshold is related to the largest eigenvalue (see Ref.~\cite{vazquez}) is recovered if we assume that all nodes have equal traversing probability $q_{k_{i}} = q = const$. In other words, we are switching from a joint site-bond percolation to a simple site percolation with an occupation probability $q$. If $q \neq 0$ (otherwise the percolation condition cannot be satisfied), we can write the condition as 
\begin{equation}
 q \det \left[ (k_{i}-1) T_{k_{i} k_{j}} p(k_{j}|k_{i}) - \Lambda \delta_{ij} \right] \geq 0~,
\label{multi17bis}
\end{equation}
with $\Lambda = 1/q$.
Since in $q=0$ the determinant is negative, the smallest positive value of $q$ ensuring Eq.~\ref{multi17bis} to be satisfied corresponds to the largest eigenvalue $\Lambda_{max}$ of the matrix $(k_{i}-1) T_{k_{i} k_{j}} p(k_{j}|k_{i})$. 
It follows that the critical value of site occupation probability is $q = 1/\Lambda_{max}$. In the case $T_{k_{i} k_{j}}=1$, the condition gives exactly the results by V\'azquez et al.~\cite{vazquez}.  

However, the complete knowledge of the correlation matrix $p(k_{j}|k_{i})$ is very unlikely for real networks, and the analytical solution of Eq.~\ref{multi17} can be problematic also for simple artificial networks. 
In the other case, in which all nodes belong to a unique class, the single-state network is recovered and the analytical treatment coincides with that presented in the previous section. 
In the rest of the paper, all the presented computations and simulation results will be referred to this particular type of
inhomogeneous joint site-bond percolation on single-state markovian networks or the corresponding explicit formulae for site or bond percolation.
Note that, starting from Eq.~\ref{mrcgg}, with few further assumptions, other simpler versions of Molloy-Reed criterion are possible, some of which will be computed in the next section.

\subsection{Transition probability factorization, single-vertex dependence and uniform probability.}

The generalized Molloy-Reed criterion holds for two-point correlated networks (Markovian networks), in which the condition for the existence of a giant component is related to the behaviour of the transition probability function $T_{k_{i} k_{j}}$, that in principle depends on edges properties and on the functional performances of the spreading process. 
Henceforth, we will focus on the problem of how different forms of transition probability
affect the inhomogeneous percolation criterion. For clarity, only the case of all vertices belonging to a single class is considered.

Although a plethora of behaviours for $T_{k_{i} k_{j}}$ could be considered, the most natural assumption consists of their factorization in two single-vertex contributions.  
Let us suppose that the transition probability $T_{k_{i} k_{j}}$ can be written in the form 
\begin{equation}
T_{k_{i} k_{j}} = \Theta_{i}(k_{i}) \Theta_{f}(k_{j})~, 
\label{teta}
\end{equation}
where subscripts $i$ and $f$ indicate an initial and a final term respectively, in order to stress the fact that first and second vertices of an edge can give different contributions in the inhomogeneous percolation process.  

Inserting Eq.~\ref{teta} in Eq.~\ref{mrcgg}, the condition for the existence of a giant component becomes
\begin{equation}
\sum_{k_{i}, k_{j} = 1}^{\infty} p_{k_{i}} k_{i} \left[ (k_{i} -1) \Theta_{i}(k_{i}) \Theta_{f}(k_{j}) q_{k_{i}} - 1 \right] p(k_{j}| k_{i}) \geq 0~. 
\label{tetateta}
\end{equation}

In the case of uncorrelated graphs, the conditional probability also factorizes in $p(k'|k) = \frac{k' p_{k'}}{\langle k \rangle}$, the generalized Molloy-Reed criterion becomes
\begin{equation}
\sum_{k_{i}, k_{j} = 1}^{\infty} p_{k_{i}} p_{k_{j}} k_{i} k_{j} \left[(k_{i} -1) q_{k_{i}} T_{k_{i} k_{j}} - 1 \right] \geq 0~,
\label{mrcggu}
\end{equation}
and Eq.~\ref{tetateta} gets simpler, leading to an interesting expression for the site percolation threshold, 
\begin{equation}
q_{c} = \frac{{\langle k \rangle}^{2}}{[ \langle k^{2} \Theta_{i}(k) \rangle - \langle k \Theta_{i}(k) \rangle ] \langle k \Theta_{f}(k) \rangle }~.
\label{tetasoil}
\end{equation}
 
As an alternative, it seems interesting to study situations in which the transition probability is a function of only
one of the two extremities of an edge. 
The case in which it depends only on final nodes means that $\Theta_{i}(k) = const. = 1$ and $T_{k_{i} k_{j}} = \Theta_{f}(k_{j})$ and the  site percolation threshold takes the form 
\begin{equation}
q_{c} = \frac{{\langle k \rangle}^{2}}{[ \langle k^{2} \rangle - \langle k \rangle ] \langle k \Theta_{f}(k) \rangle}
= q_{c}^{hom}\frac{\langle k \rangle}{\langle k \Theta_{f}(k) \rangle}~,
\label{tetasoil2}
\end{equation}
where $q_{c}^{hom}$ is the value of the critical occupation probability for the correspondent standard homogeneous site percolation.
The opposite situation, $\Theta_{f}(k) = const. = 1$ and $\Theta_{i}(k)= T_{k}$, leads to the following form for the Molloy-Reed criterion
 \begin{equation}
\sum_{k=1}^{\infty} k \left[(k - 1)q_{k}T_{k} - 1 \right] p_{k}  \geq 0~. 
\label{eq_1k}
\end{equation} 
The inhomogeneous site percolation threshold $q_{c}$ follows directly imposing uniform occupation probability $q_{k}=q$,
\begin{equation}
q_{c} = \frac{\langle k \rangle}{\langle k^{2} T_{k} \rangle - \langle k T_{k} \rangle}~. 
\label{eq_2k}
\end{equation}
From Eq.~\ref{eq_1k}, also the threshold's value for the inhomogeneous bond percolation is immediately recovered. 
If we suppose uniform transition probabilities $T_{k} =T$, indeed, we can explicit $T$ in Eq.~\ref{eq_1k} as a function of the set of $\{q_{k}\}$ and compute the value of the threshold as
\begin{equation}
T_{c} = \frac{\langle k \rangle}{\langle k^{2} q_{k} \rangle - \langle k q_{k} \rangle}~, 
\label{eq_3k}
\end{equation}
that represents the case in which percolation on the edges is affected by refractory nodes. 
Such a situation can have applications in the study of real networks, expecially in relation to edge removing procedures for estimating networks resilience but, in the rest of this paper, we limit our analysis to the site percolation. 

The last case we consider is that of uniform transition probabilities $T_{k_i k_j} = T$, with $0 < T \leq 1$.
In the limit $T = 1$, the original Molloy-Reed criterion is recovered, otherwise, introducing $T_{k_i k_j} = T$ in  Eq.~\ref{mrcggu}, we find the same expression for the threshold of site percolation with noisy edges (as in Eq.~\ref{thres0} with $\langle T \rangle = T$)
\begin{equation}
q_{c} = \frac{1}{(\langle k^{2}\rangle/\langle k \rangle - 1) T}~. 
\label{eq_1}
\end{equation} 

~\\

\section{APPLICATIONS AND NUMERICAL SIMULATIONS}
\label{appsec}

In Section~\ref{intro}, it has been stressed that the inhomogeneous percolation could be exploited to investigate the relation between  structural and functional properties of networks. For example, compared to standard percolation, the inhomogeneous one provides information on the {\em functional robustness} of a graph, that is the feedback of the graph to the presence  of dynamical processes spreading on it. The topological robustness of a graph is principally studied looking at the size of the giant component, assuming that connectivity properties are sufficient to discriminate between ``robust'' and ``fragile'' graphs. In real graphs, as communication networks, our perception of the robustness is based on the observation of flows of information and other physical quantities: a connected component can be considered robust if a flow can span throughout it without stopping or causing any failure  (e.g. in the transmission of an amount of data or energy). 
Since percolation properties are used for predicting the robustness or the vulnerability of a graph, in the following we will often translate percolation results to the more effective language of network's resilience, exploiting the equivalence between the percolation threshold and the critical fraction $f_{c}$ of removed vertices. 
As mentioned in Section~\ref{intro}, the robustness is computed removing a finite fraction $f$ of vertices (and the edges connected to them) and studying the size $\mathcal{G}_{f}$ of the giant component as a function of $f$.
The removal of a node corresponds exactly to a unoccupied node in the percolation model, then if the percolation threshold $q_{c}$ exists, the relation $f_c = 1- q_c$ holds between it and the threshold value of the fraction of removed nodes above which the graph does not admit a giant component.
Since the presence of edge transition probabilities affects the value $f_{c}$ and the overall robustness of the network, decreasing the size of the giant component, it suggests that inhomogeneous percolation is a good paradigm for the study
of the functional robustness. 

The functional dependence of the transition probabilities $\{T\}$ on the vertices and the edges is strongly related to the specifics of the system and to the type of processes one is interested to describe. Therefore, in the absence of hints from studies of real data about spreading processes on networks, the present section is devoted to the highlight of some simple examples of functional forms for the transition probability, and to the illustration of their effects on the percolation condition. We limit our analysis to two main classes of random graphs: the homogeneous graphs, in which the connectivity distribution is peaked around a characteristic degree value, and the heterogeneous graphs, whose degree distribution is broad, with very large fluctuations in the degree values. Typical examples of this second class are scale-free random graphs with power-law degree distributions ($p_{k} = {\zeta(\gamma)}^{-1} k^{-\gamma}$), while for homogeneous graphs we will always consider poissonian degree distributions ($p_{k} = e^{-\langle k \rangle} {\langle k \rangle}^{k}/k!$). We also present results of numerical simulations validating our analytical findings.

The numerical simulations have been performed using efficient algorithms for percolation on graphs. In particular, the algorithms based on a depth-first search are those better encoding the idea of a spreading process that starting from a node pervades the network. For this reason, we preferred to use this type of algorithm. Moreover, agglomerative tree-based 
algorithms~\cite{ziff} can be correctly used only if edge transition probabilities are symmetric, otherwise we cannot be sure that the implicit direction of the spreading is conserved during the agglomeration process of the discovered rooted trees.   
Hence, after having randomly chosen a starting node, we perform a depth-first search and compute the size of the connected component the node belongs to. This operation is repeated for all the nodes of the network (excluding those already visited): the size of the giant component is recorded. Averaging over many realizations of networks, we get a statistical estimation of the size of the giant component for a networks ensemble with a given set of properties.
 
\subsection{How different forms of transition probability affect percolation properties}

In the following, we analyze the effects that different degree-dependent edge transition probabilities can have on the functioning of homogeneous and heterogeneous graphs, i.e. how they affect the functional giant component of these graphs.
In particular, we take into account examples from three different situations: the case in which the transition probability  depends on two factorized functions of the degrees at the extremities of the edge, the case of single-vertex transition probability monotonically increasing or decreasing with $k$ and the case of uniform or randomly distributed transition probability. Note that, for the sake of simplicity, in the following we always consider uncorrelated networks.
 
\subsubsection{Factorized degree-dependent transition probability}
Let us begin with a transition probability that increases with the degree: since the transition probability takes values in the interval $[0,1]$, we assume it tends to $1$ for edges joining highly connected nodes, that is reasonable in dynamical processes such as the exchange of information on the Internet. For example, in communication networks, the hubs have very powerful and efficient systems to handle and re-direct information to the periphery and their mutual connections possess huge bandwidths compared to single user's connections. 

As an example of this situation, we consider $T_{k k'}$ to be a generic homogenous function in $k$ and $k'$ (because the edges are undirected) and assume that it depends only on the product $kk'$ of the degrees of the edge extremities.
This is justified by the fact that, in many weighted networks, the transition probability can reasonably be a function of edge weights $T_{ij}=T(w_{ij})$, since weights are usually measures of the flow and the traffic on the edges, or estimations of edges capacity.
In this regard, a recent study by Barrat et al. \cite{barrat} on the architecture of weighted networks, analyzes the correlations between topological and weighted quantities in the real weighted network of world-wide air-transportations, providing an indirect justification for our choice. In fact, it turns out that edge weights are correlated to the degrees at their extremities according to a phenomenological law of the type $w_{ij} \sim {(k_{i} k_{j})}^{\alpha}$, $\alpha >0$ ($\alpha \sim 0.5$ for the airports network).
More recently, other papers on weighted graph modeling \cite{serrano,almaas} have used this relation as
a reasonable form for the correlation between weights and degrees.

%
%
\begin{figure} 
\centerline{
\includegraphics*[angle=-90,width=0.5\textwidth]{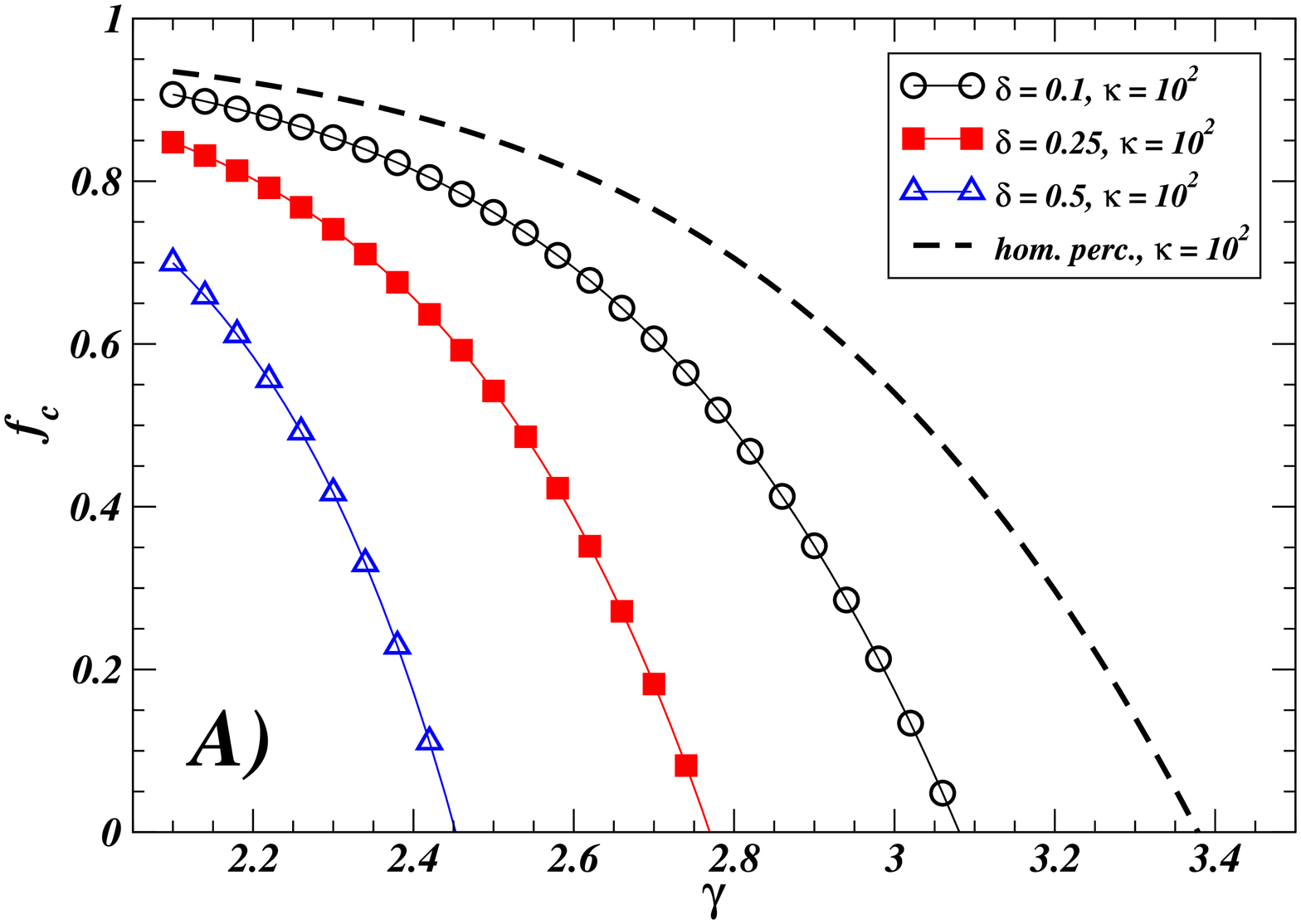}~~
\includegraphics*[angle=-90,width=0.5\textwidth]{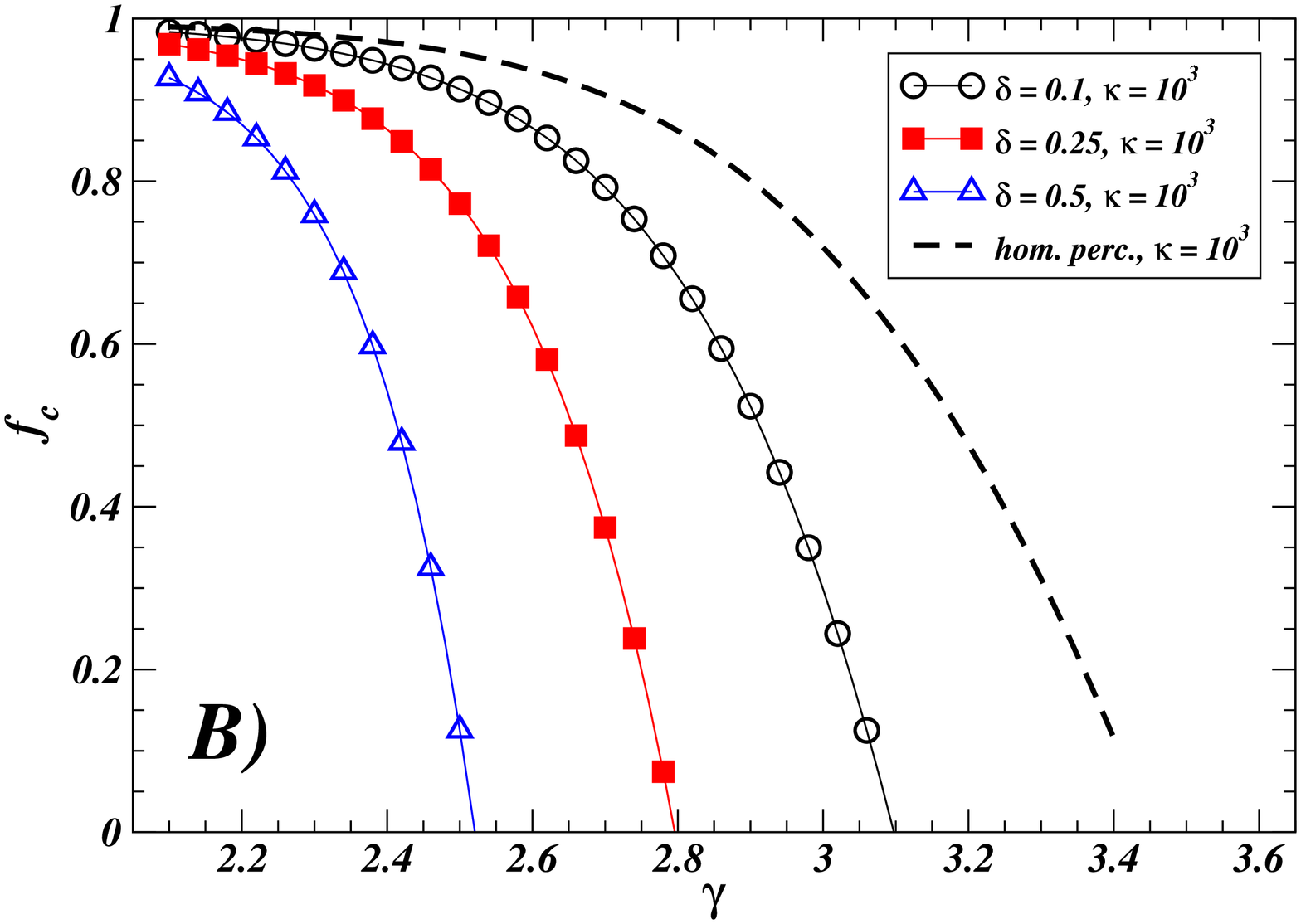}
}
\caption{The two figures show the behaviour of $f_{c} = 1- q_{c}$ from Eq.~\ref{qgamdel} for a power-law random graph with cut-off $\kappa= 10^2$ (A) and $10^3$ (B).  The fraction $f_{c}$ of removed nodes required to destroy the giant component $\mathcal{G}_{f}$ is plotted as a function of the exponent $\gamma$ of the power-law when the transition probability reads $T_{k k'} \propto {(k k' /{\kappa}^2)}^{\delta}$,  with exponent $0 < \delta <1$. Three values of $\delta$ are explored: $\delta=0.1$ (circle), $\delta=0.25$ (squares) and $\delta=0.5$ (triangles). Dashed lines show the solution for standard homogeneous site percolation (limit of $\delta \rightarrow 0$)\\}
\label{figura1}
\end{figure}
%
%
%
%
\begin{figure} 
\begin{center}
\includegraphics[angle=-90,width=0.5\textwidth]{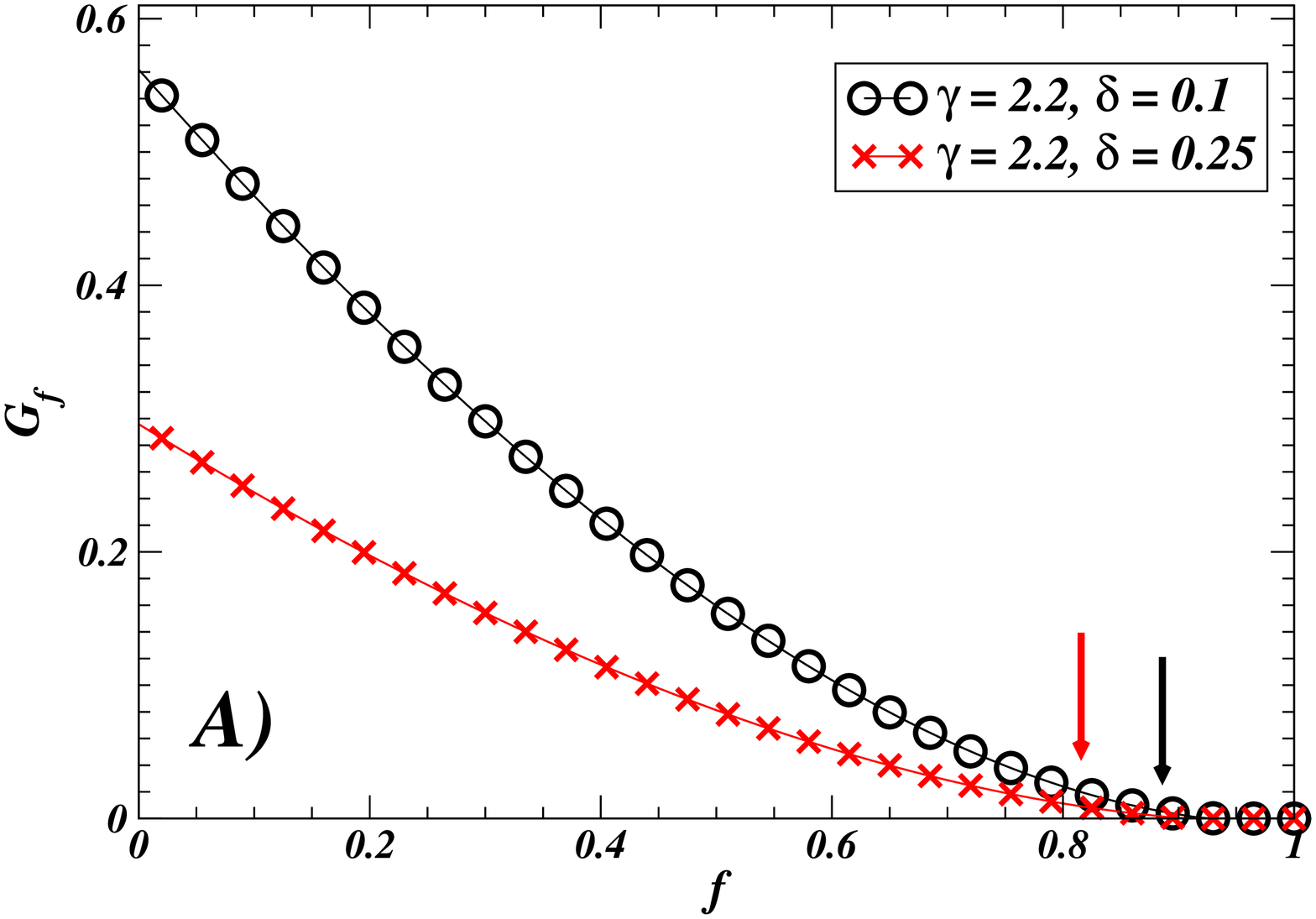}~
\includegraphics[angle=-90,width=0.5\textwidth]{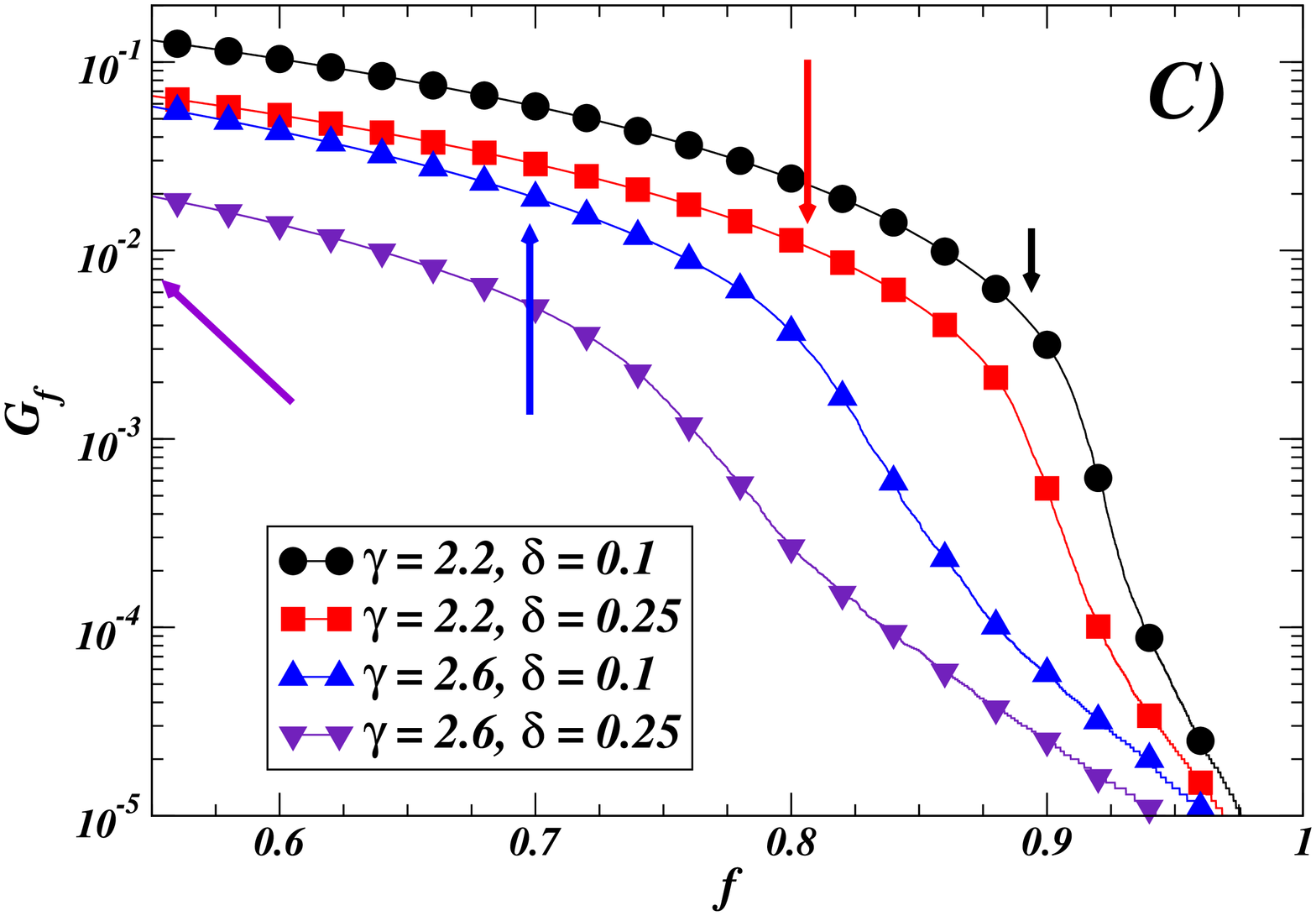}\\~\\
\includegraphics[angle=-90,width=0.5\textwidth]{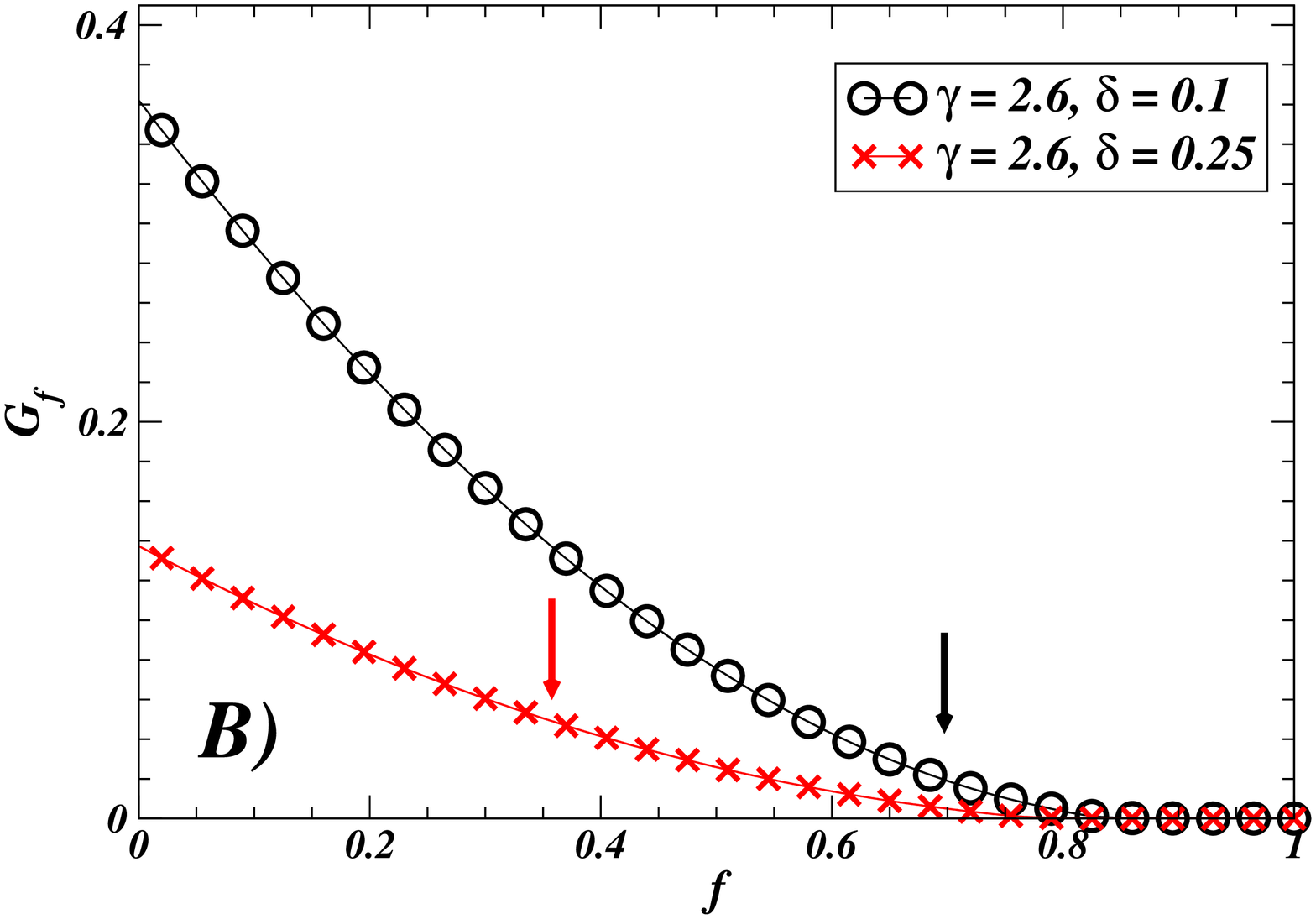}~ 
\includegraphics[angle=-90,width=0.5\textwidth]{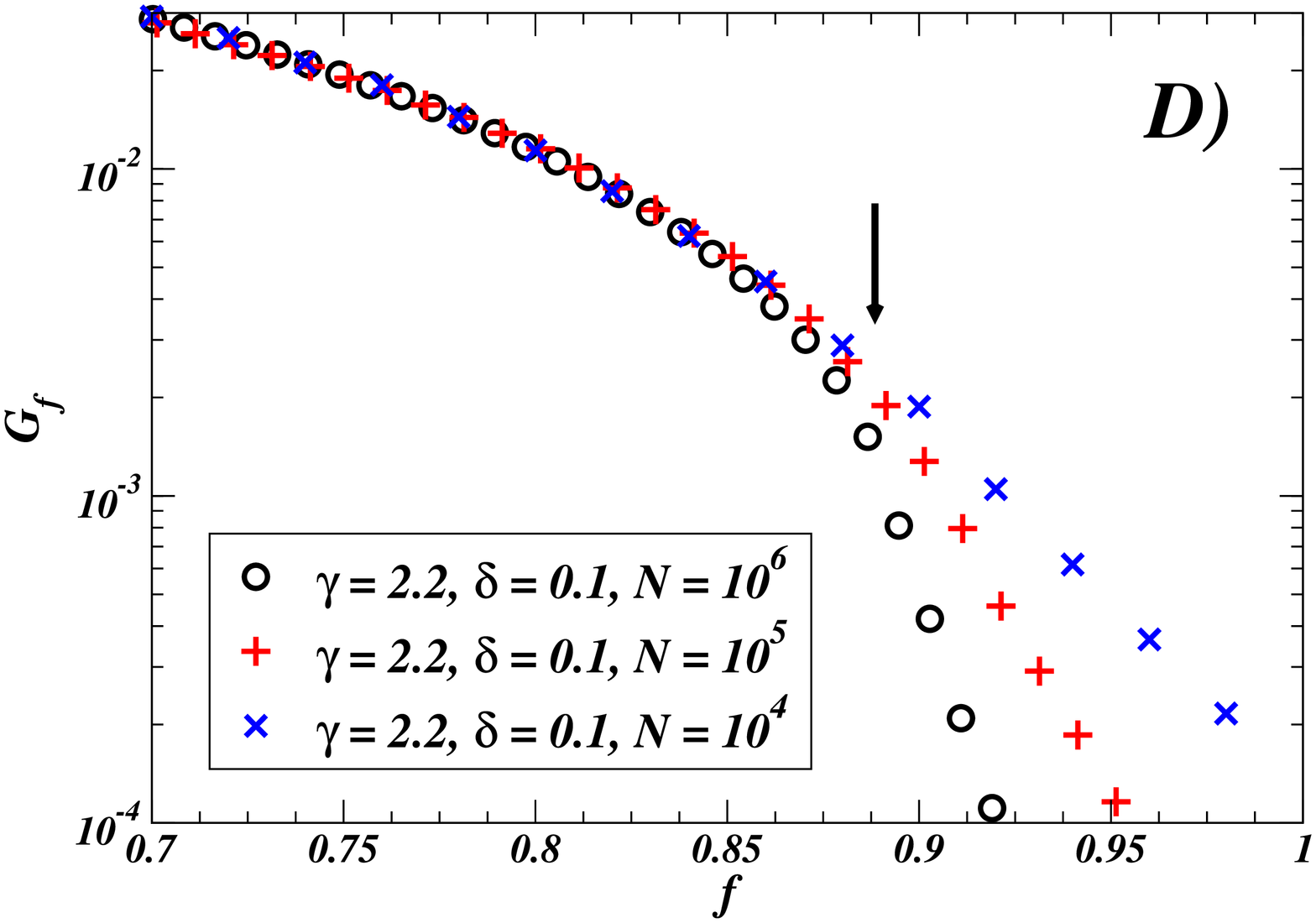} 
\end{center}
\caption{A-B) The figures show the behaviour of the giant component $\mathcal{G}_{f}$ as a function of the fraction $f$ of nodes removed for a power-law graph with $N=10^6$ nodes (averaged over $100$ realizations), exponent $\gamma =2.2$ (A) and $\gamma = 2.6$ (B), and with edge transition probabilities $T_{k_{i} k_{j}} = ({\frac{k_{i} k_{j}}{{\kappa}^2})}^{\delta}$.
Two values of $\delta$ are explored: $\delta = 0.1$ (circles) and $\delta=0.25$ (crosses). The cut-off is $\kappa = 10^2$. The arrows point to the threshold's position predicted by Eq.~\ref{qgamdel}.\\ 
C-D) The two figures report data on the numerical accuracy of the analysis. Figure C displays a lin-log scale plot of curves in panels A and B, revealing that the determination of the percolation treshold is affected by the bad convergence of the curves, in particular for large values of $\gamma$. Figure D shows the strong finite-size effects on the data; we consider samples with identical parameters ($\gamma = 2.2$, $\delta = 0.1$ and $\kappa = 10^2$) and different graph's size $N=10^4$ (crosses), $10^5$ (pluses)and $10^6$ (circles). The arrows indicate the positions of the threshold predicted by Eq.~\ref{qgamdel}. \\}
\label{simul1}
\end{figure}
%
%
According to this observation, we define transition probabilities $T_{k k'} = T_{1} {(\frac{k k'}{{\kappa}^{2}})}^{\delta}$ with $\delta >0$ and cut-off $\kappa$. By computing the threshold for site percolation with inhomogeneous edges directly from Eq.~\ref{thresh1}, we get
\begin{equation}
q_{c}(\gamma, \delta)  = \frac{{\langle k \rangle}^{2} {\kappa}^{2\delta}}{T_{1}(\langle {k}^{2+\delta} {k'}^{1+\delta} \rangle - \langle {k}^{1+\delta} {k'}^{1+\delta} \rangle)} = \frac{{\langle k \rangle}^{2} {\kappa}^{2\delta}}{T_{1} \langle k^{1+\delta} \rangle [\langle k^{2+\delta} \rangle - \langle k^{1+\delta} \rangle]}~.
\label{qgamdel}
\end{equation}
Particularly interesting is the behaviour of $q_{c}$ for power-law graphs with degree distribution $p_{k}= {\zeta(\gamma)}^{-1} k^{-\gamma}$, that we have computed by numerical summation. We also put $T_{1} = 1$.
In Fig.~\ref{figura1}, we report two sets of numerical curves giving $f_{c} = 1-q_{c}$ as function of the topological exponent $\gamma$ for fixed values of $\delta$ and of the cut-off $\kappa$: panel A displays the curves for $\kappa = 10^2$ and $\delta=0.1$ (circles),  $0.25$ (squares), $0.5$ (triangles), while those for $\kappa=10^3$ at the same values of $\delta$ are shown in panel B. 
Dashed lines show the behaviour of the threshold value for the standard site percolation ($\delta \rightarrow 0$). 
The plots show clearly that the fraction of removed nodes sufficient to disconnect the giant component is considerably reduced when such a transition probability is taken into account and that the effect is emphasized by the cut-off.
The results of simulations on power-law graphs with up to $N=10^6$ vertices are reported in Fig.~\ref{simul1}. Panels A-B show typical experimental curves for the giant component $G_{f}$ as a function of the removed fraction of vertices $f$ for two different exponents $\gamma =2.2$ (A), and $\gamma = 2.6$ (B), cut-off value $\kappa = 10^2$ and two different values of $\delta = 0.1,$ and $\delta=0.25$. The curves for $\gamma = 2.2$ give a threshold estimation that is in good agreement with the predicted values. In the case of larger exponent, on the contrary, the agreement gets worse with a clear overestimation of the percolation threshold. The criterion itself for numerically detecting the threshold value (i.e. a flex point in the shape of the curves) becomes somewhat arbitrary as shown by the lin-log plots of the curves in panel C. However, we have checked these data on different sizes and plotted the corresponding results in panel D of Fig.~\ref{simul1}. The three curves represent numerical values of the giant component $\mathcal{G}_{f}$ as a function of the removed fraction of nodes for $N=10^4$ (crosses), $N=10^5$ (pluses) and $N=10^6$ (circles). The curves become steeper and steeper near the threshold as the size increases, meaning that only for very large graphs the theoretical values can be reached.     
Finally, the agreement between simulations and theoretical predictions are slightly better if we put a larger cut-off, for example $\kappa = {10}^{3}$.

%
%
\begin{figure} 
\centerline{
\includegraphics*[angle=-90,width=0.5\textwidth]{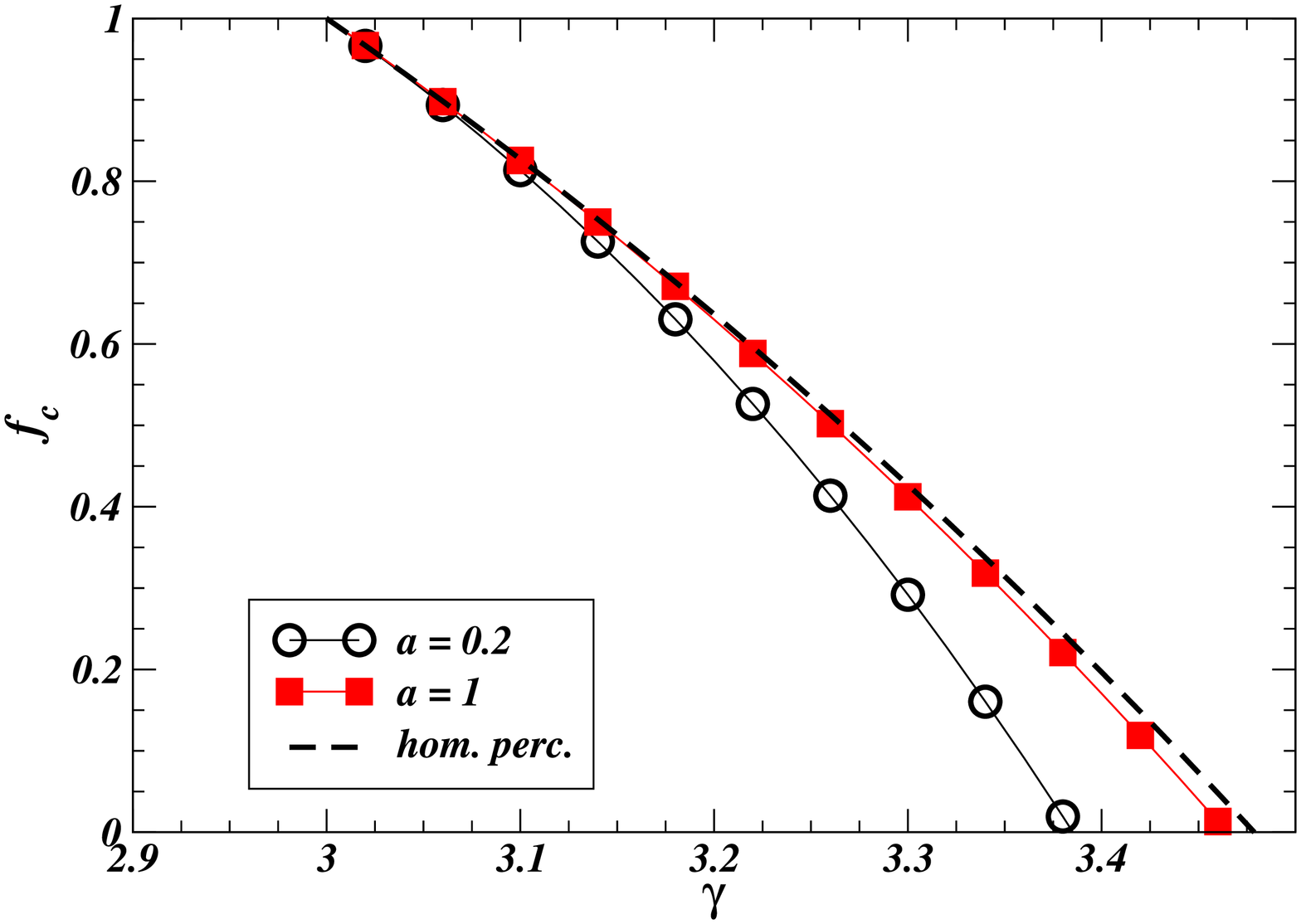}
}
\caption{Critical fraction $f_{c}$ of removed nodes required to destroy the giant component $\mathcal{G}_{f}$ as a function of the exponent $\gamma$ for a power-law graph in which the transition probability function is single-vertex depending with the form $T_{k} \sim 1- \exp(-\frac{a k}{\langle k \rangle})$. The figure reports the behaviour of the expression $f_{c} = 1-q_{c}$ with $q_{c}$ computed from Eq.~\ref{q_expa} and correct for an infinite graph. The parameter $a$ controls the saturation to $1$ for highly connected nodes: the cases $a=0.2$ (circles) and $a=1$ (squares) are shown. For large values of the parameter $a$, the curves rapidly converge to the limit behaviour of homogeneous percolation (limit $a \rightarrow \infty$). For $a <1$, the curves shift from this limit, but in the range $2 < \gamma <3$ diverging fluctuations ensure 
its robustness (and the existence of percolation).\\}
\label{figura2}
\end{figure}
%
%

\subsubsection{Single-vertex dependent transition probability}
Another interesting case is that of the transition probability depending only on the sender's properties, i.e. only on the degree of the initial vertex. 
A first example is that of a monotonously increasing functional form of the degree, that saturates to $1$ at large $k$ values: we consider $T_{i j} = T_{k_{i}} = \Theta_{i}(k) \sim 1-\exp(-{\frac{a k}{\langle k \rangle}})$, that can be applied also to infinite graphs. The parameter $a$ controls the convergence to $1$.
For large scale-free networks, however, the mere presence of a saturation, rather than its rapidity, is sufficient for  
the system to behave as in the homogeneous percolation, showing that graph's heterogeneity ensures a zero percolation threshold.
From Eq.~\ref{eq_2k} simple calculations lead to the expression of the threshold for site percolation with inhomogeneous edges,
\begin{equation}
q_{c} = \frac{\zeta(\gamma-1)}{\zeta(\gamma-2) - \zeta(\gamma-1) + {Li}_{\gamma-1}(e^{-a\zeta(\gamma)/\zeta(\gamma-1)}) - {Li}_{\gamma-2}(e^{-a\zeta(\gamma)/\zeta(\gamma-1)})}~,
\label{q_expa}
\end{equation}
in which $Li_{\gamma}(z)$ is the polylogarithmic function $\sum_{k=1}^{\infty} \frac{z^{k}}{k^{\gamma}}$ \cite{stegun}.  
In Fig.~\ref{figura2}, the behaviours of $q_{c}$ vs. $\gamma$ for some values of the parameter $a$ are reported. They show that the critical fraction of removed nodes $f_{c}$ is exactly $1$ for $2 < \gamma < 3$. Also for $\gamma \geq 3$, the behaviour is qualitatively the same as for standard homogeneous percolation (and corresponding robustness measures) on scale-free networks.
The reason is essentially that a transition probability converging to $1$ for the highly connected nodes does not affect the network's properties if the network is infinite, because there is always a considerable fraction of nodes with optimal transmission capability.
This condition is not trivially satisfied by finite graphs, and the presence of a cut-off in the degree can have a strong influence on the network's functional robustness.
Two kinds of cut-offs on power-laws have been largely studied in literature: an abrupt truncation of the degree distribution at a maximum value $\kappa \approx N^{\frac{1}{\gamma-1}}$, or a natural exponential cut-off, i.e. $p_{k} \sim k^{-\gamma} e^{-k/\kappa}$.
%
\begin{figure} 
\centerline{
\includegraphics*[angle=-90,width=0.5\textwidth]{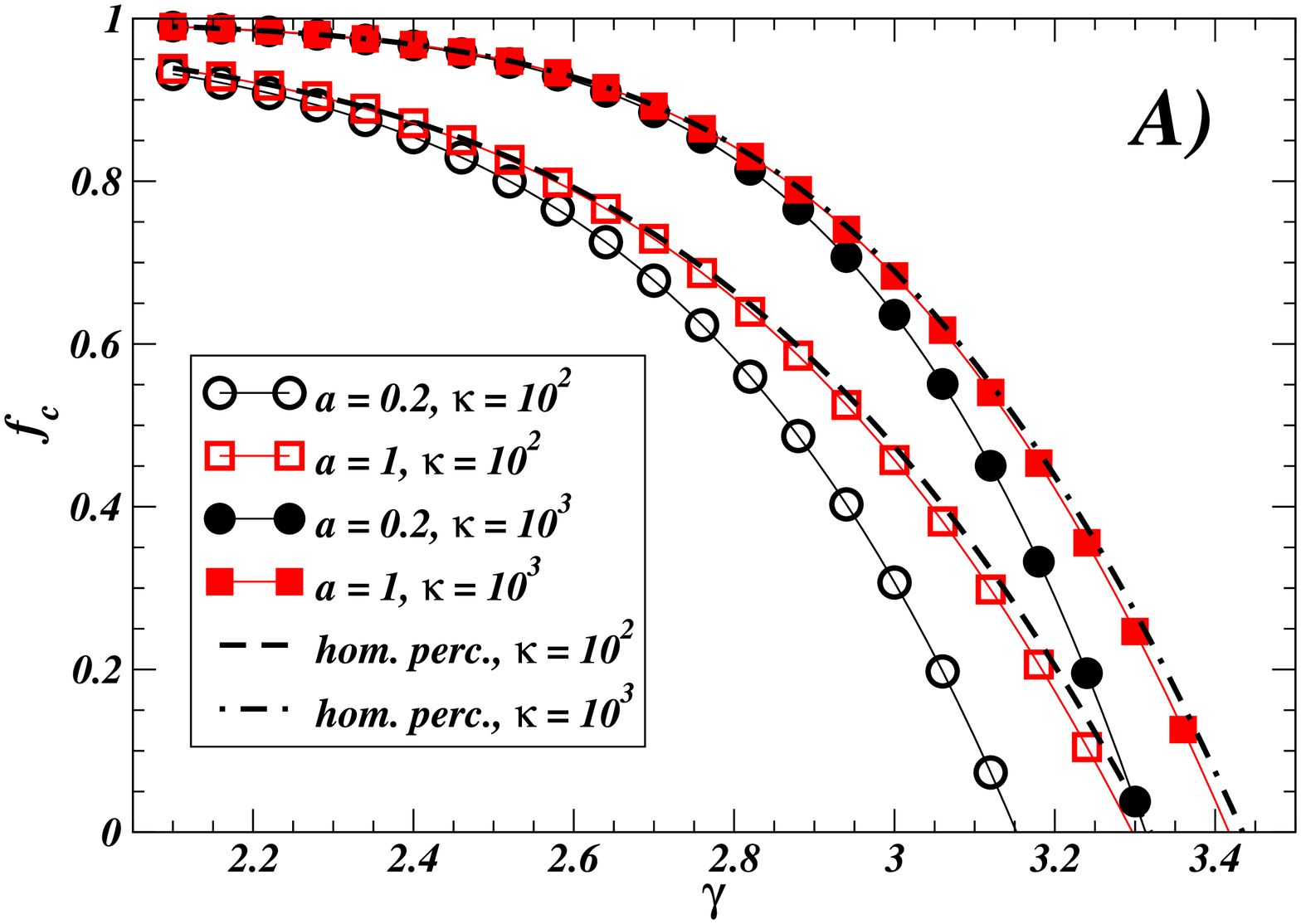}~~
\includegraphics*[angle=-90,width=0.5\textwidth]{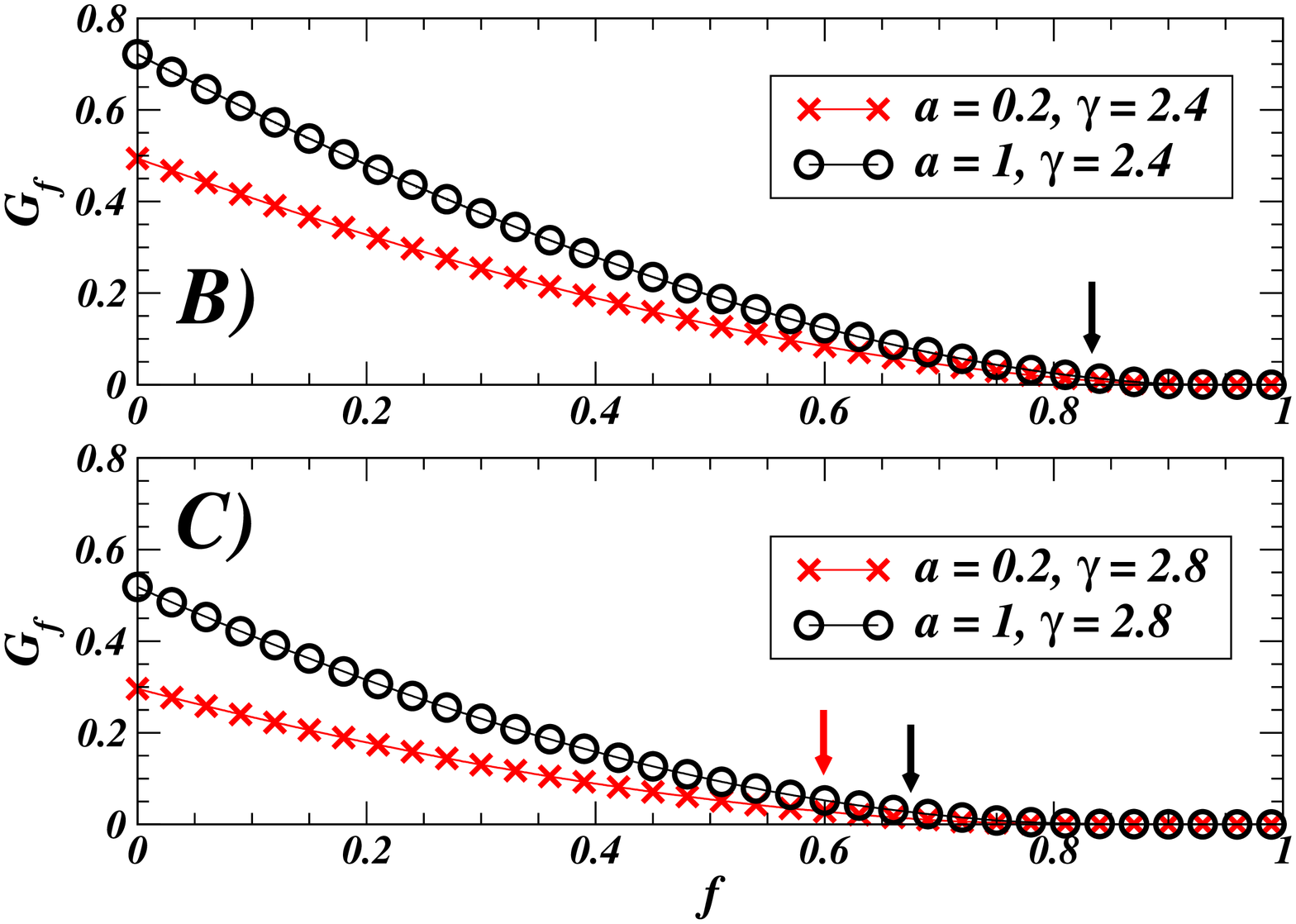}
}
\caption{
A) The fraction $f_{c}$ of removed nodes required to destroy the giant component $\mathcal{G}_{f}$ as a function of the exponent $\gamma$ for a power-law graph when the transition probability function is single-vertex depending $T_{k} \sim 1- \exp(-\frac{a k}{\langle k \rangle})$. The parameter $a$ controls the saturation to $1$ for highly connected nodes. The figure reports the numerical computation of the expression $f_{c} = 1-q_{c}$ with $q_{c}$ from Eq.~\ref{q_expa} when, in the infinite graph, we put a cut-off on the degree. The cut-off values are $\kappa = 10^2$ (open symbols) and $10^3$ (full symbols) and the control parameter is $a= 0.2$ (circles) and $1$ (squares). The critical fraction is lower than $1$ in all the range $2 < \gamma <3$, but in agreement with the results for standard homogeneous percolation (dashed and dotted-dashed lines).\\
B-C) The curves $\mathcal{G}_{f}(f)$ obtained by numerical simulations on power-law random graphs of size $N=10^5$ and exponent $\gamma = 2.4$ (B) and $\gamma = 2.8$ (C) for the two values of $a = 0.2, 1$ and cut-off $\kappa = 10^2$. The arrows indicate the positions of the threshold predicted using Eq.~\ref{qLi}.
\\}
\label{figura3}
\end{figure}
%
%

While finite-size effects can be studied only by numerical summation, the expression for a degree distribution with exponential cut-off can be solved in a closed form introducing the corresponding $p_{k}$ in Eq.~\ref{eq_2k}: the critical percolation threshold is readily expressed using polylogarithmic functions as
\begin{equation}
q_{c}= \frac{{Li}_{\gamma-1}(e^{-\frac{1}{\kappa}})}{{Li}_{\gamma-2}(e^{-\frac{1}{\kappa}})-{Li}_{\gamma-1}(e^{-\frac{1}{\kappa}})+{Li}_{\gamma-1}(e^{-\ell(\gamma, \kappa)})-{Li}_{\gamma-2}(e^{-\ell(\gamma, \kappa)})}~,
\label{qLi}
\end{equation}
with $\ell(\gamma, \kappa)$ defined by
\begin{equation}
\ell(\gamma, \kappa)= \frac{1}{\kappa} + \frac{a {Li}_{\gamma}(e^{-\frac{1}{\kappa}})}{{Li}_{\gamma-1}(e^{-\frac{1}{\kappa}})}~.
\end{equation}
The introduction of a cut-off actually reduces the effects of degree fluctuations, producing finite percolation threshold also in the range $2 < \gamma <3$.  
The two methods of fixing the cut-off give qualitatively similar results about the decreasing behaviour of robustness (data not shown), even if natural exponential cut-offs produce slightly lower values. 
However, inhomogeneous percolation does not seem to enrich the scenario obtained by standard percolation, though, the case of slow convergence ($a <1$) yields a reduction of the functional robustness.  
Abrupt truncation allows to consider examples that are equivalent to numerical simulations on computer-generated graphs and to compare them. 
In Fig.~\ref{figura3}-A two groups of curves are reported, showing data of the numerical computation of the critical fraction $f_{c}$ for a power-law graph with different values of the cut-off $\kappa= 10^2, 10^3$ as a function of the exponent $\gamma$. 
Numerical simulations for power-law graphs with $N={10}^5$ vertices (and $100$ realizations) and cut-off $\kappa = {10}^{2}$ are reported in Fig.~\ref{figura3} (panels B-C). The curves in panel B refer to the case of $\gamma = 2.4$ and $a= 0.2, 1$, they are in good agreement with the threshold values predicted by the theory. In panel C, the two curves refer to graphs with exponent $\gamma = 2.8$ and the same values of the parameter $a$. This time the actual collapse of the curves takes place at larger values of $f$ if compared with theoretical predictions. For $a=0.2$ and $\gamma=2.8$ the threshold value should be close to $f=0.6$, while numerical simulations give at least $0.7$. 

However, the estimation of the threshold value, as also shown in Fig.~\ref{simul1}, is a very difficult task in power-law graphs, given that the shape of the curves decreases with constant convexity, converging very smoothly to zero. As we will see in the next example, simulations on homogeneous graphs provides considerably better results. A possible explaination for the occurrence of these biases comes from the fact that the probability of nodes with maximum degree ($k \approx \kappa$) is $\mathcal{O}(1/N)$, then in a finite graph their frequency depends on the size of the system.  Compared to the homogeneous case, the traffic on heterogeneous graphs is clearly unbalanced in favor of large degree nodes, therefore we argue that finite size effects can be more relevant. Similar effects are visible also in the case of optimal transmission ($T=1$ for all edges), albeit they are emphasized when the global efficiency of the network is decreased. 
Moreover, this argument agrees with the progressive steepening of the curves for larger sizes as shown in panel D of Fig.~\ref{simul1}.        

%
\begin{figure} 
\centerline{
\includegraphics*[angle=-90,width=0.5\textwidth]{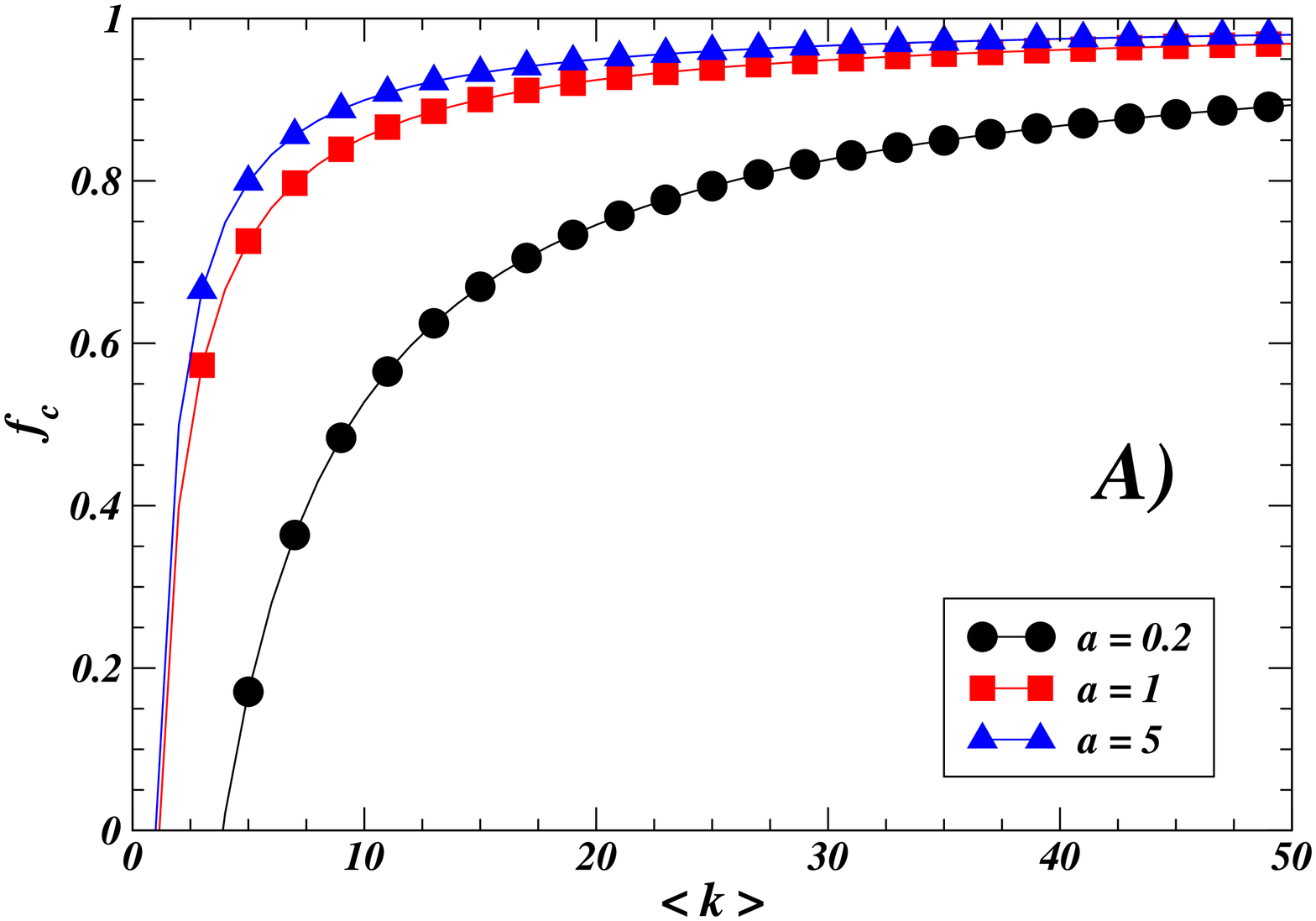}~~
\includegraphics*[angle=-90,width=0.5\textwidth]{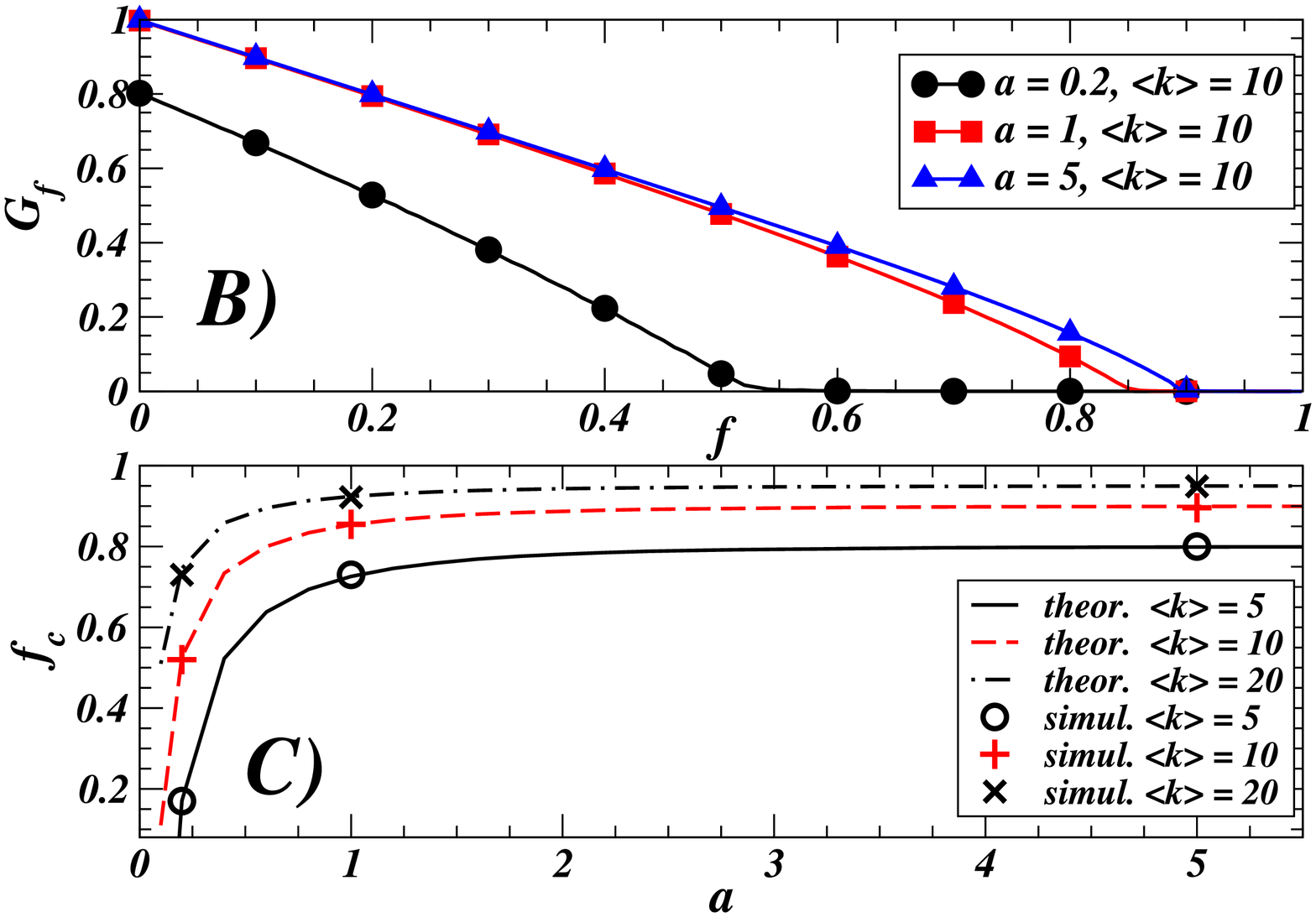}
}
\caption{
A) Fraction $f_{c}$ of removed nodes required to destroy the giant component as a function of the average degree  $\langle k \rangle$ for an Erd\"os-R\'enyi random graph with single-vertex transition probability $T_{k} \sim 1- \exp(-\frac{a k}{\langle k \rangle})$. The curves correspond to different values for the control parameter $a$: $a=0.2$ (circles), $1$ (squares) and $5$ (triangles).
Highly connected random graphs are extremely robust, but many real networks have low average connectivity, range in which
the robustness decreases considerably.\\
B-C) Panel B displays the size of the giant component $\mathcal{G}_{f}$ as a function of the removed fraction $f$ of nodes obtained by simulations on Erd\"os-R\'enyi random graphs of $N= 10^5$ nodes (averaged over $100$ realizations), average degree $\langle k \rangle = 10$ and parameter $a = 0.2$ (circles), $1$ (squares) and $5$ (triangles).
In panel C, numerical values (points, obtained by simulations) and theoretical predictions (lines) of the corresponding percolation threshold for Erd\"os-R\'enyi random graphs with different average degree $\langle k \rangle = 5, 10, 20$. 
\\}
\label{figura4}
\end{figure}
%
%

Structural properties of homogeneous graphs, on the contrary, are always deeply affected by this type of transition probability, expecially when perfect transmission is reached beyond the peak of the degree distribution ($a < 1$). 
The threshold for the site percolation with inhomogeneous edges is a function of the average degree $z = \langle k \rangle$, with the parameter $a$ adjusting saturation's rapidity,
\begin{equation}
q_{c}(a,z)= \frac{1}{z [1- \exp[\frac{{z}^{2} (\exp(-\frac{a}{z})-1) - 2a}{z}]]}~.
\end{equation}
We have plotted in Fig.~\ref{figura4}-A the corresponding curves for the critical fraction of removed nodes $f_c$ determining network's collapse.  
The results of numerical simulations on Erd\"os-R\'enyi random graph with $N=10^5$ vertices and $\langle k \rangle =10$ are also shown in Fig.~\ref{figura4} (panels B-C). The three curves in panel B represent the size of the giant component $\mathcal{G}_{f}$ as a function of $f$ for different values of the parameter $a=0.2$ (circle), $a=1$ (squares) and $a=5$ (triangles). The figure C shows the good agreement between the numerical values of the percolation threshold (points) obtained by simulations and the correspondent theoretical predictions (lines) for different values of the average degree $\langle k \rangle = 5, 10, 20$.

%
\begin{figure} 
\centerline{
\includegraphics*[angle=-90,width=0.5\textwidth]{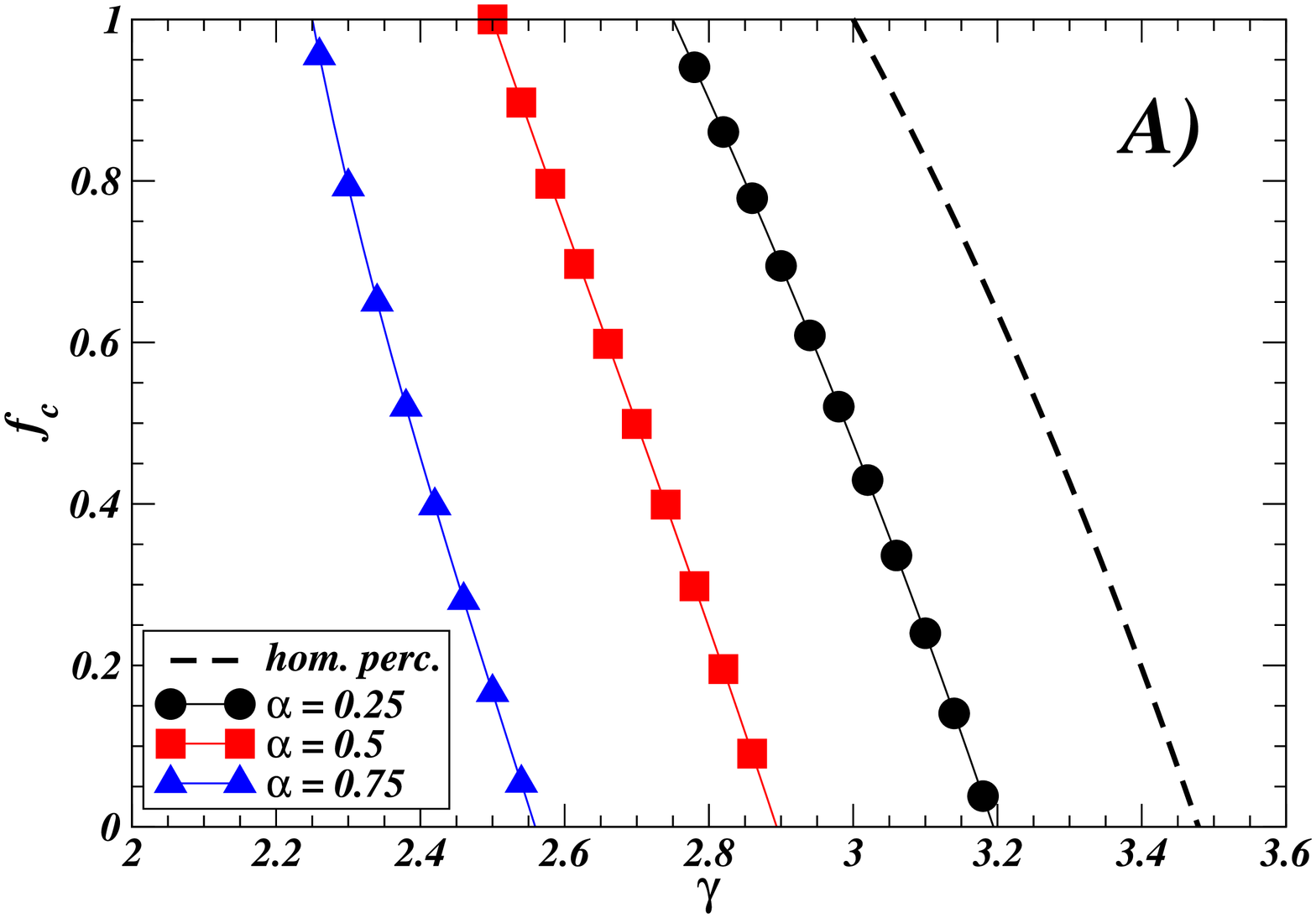}~~
\includegraphics*[angle=-90,width=0.5\textwidth]{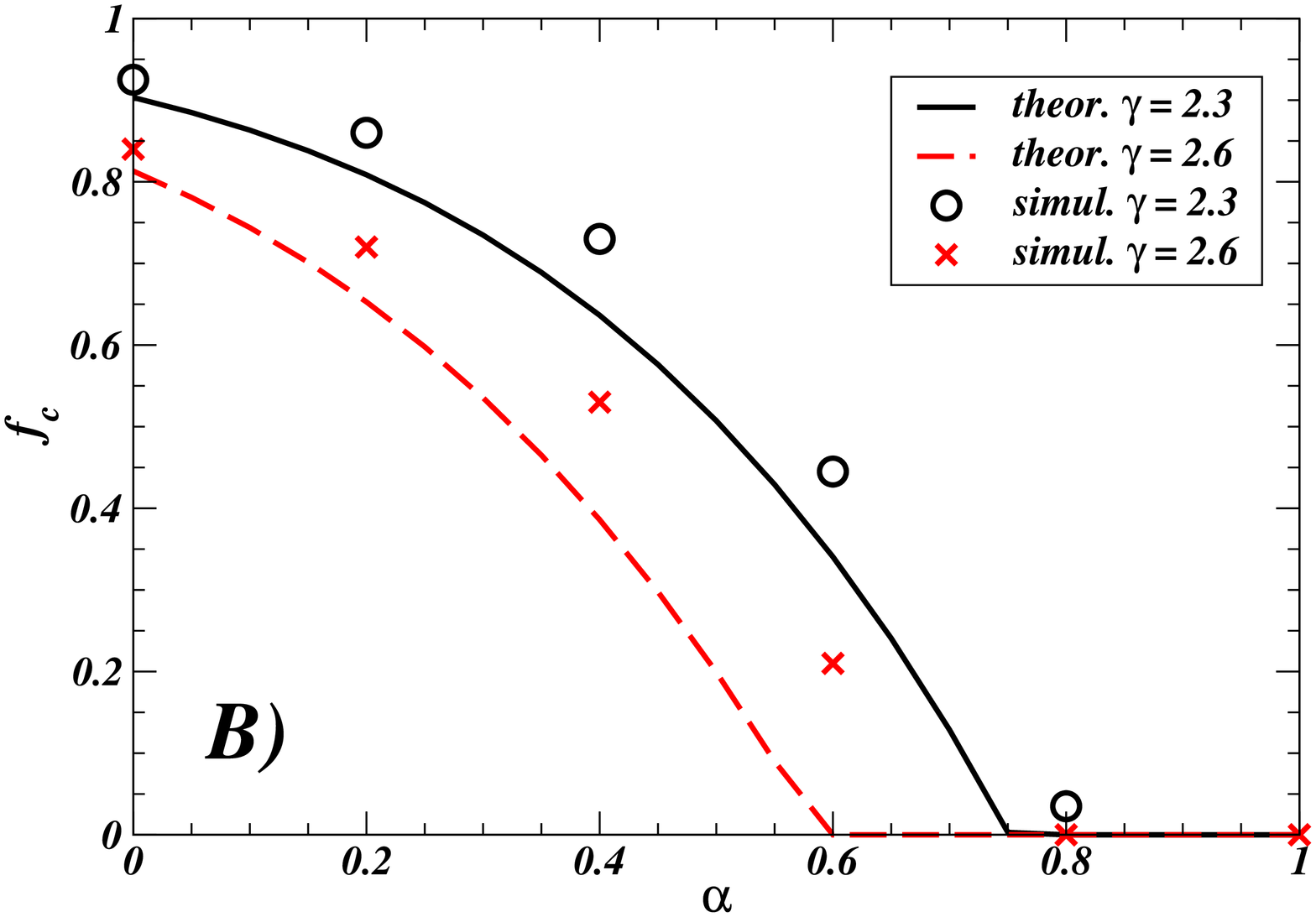}
}
\caption{A) Fraction $f_{c}$ of removed nodes required to destroy the giant component as a function of the exponent $\gamma$ for a power-law graph in which the transition probability is the single-vertex function $T_{k} \sim {k}^{-\alpha}$,  with exponent $0 < \alpha <1$. The curves are report the values of $f_{c}=1-q_{c}$ with $q_{c}$ computed by Eq.~\ref{q_alp}. All curves for $\alpha = 0.25$ (circles), $0.5$ (squares) and $0.75$ (triangles) show that the network's robustness is highly diminished compared to the homogeneous result (dashed line), also for an infinite graph. In particular, a finite threshold lower than $1$ appears also in the range $2 < \gamma < 3$.\\
B) Threshold value $f_{c}$ of removed nodes as a function of the parameter $\alpha$ for power-law graphs with cut-off $\kappa = 10^2$ and exponent $\gamma = 2.3$ (full line and circles) and $\gamma = 2.6$ (dashed line and crosses). The symbols (circles and crosses) are results from simulations on random graphs of size $N= 10^5$ (average over $100$ realizations); the lines are the correspondent theoretical predictions.\\}
\label{figura5}
\end{figure}
%
%

~\\

Another wide class of transition probabilities contains those converging to zero with increasing degree. 
At a first glance, it seems a very unphysical condition, but the problem can be inverted asking which is the maximal decay in the degree dependence still ensuring site percolation.  
For instance, exploiting bond percolation to study epidemic spreading, Newman has shown \cite{newman1} that if transmissibility decreases as $1/k$ or faster, there are no global outbreaks. Here, we can easily recover the same result expliciting the threshold value with the more appropriate tool of inhomogeneous site percolation. 
For single-vertex transition probability $T_{k}$, Eq.~\ref{eq_2k} implies that, independently of the degree distribution, a graph does not admit percolation if $T_{k}$ decays as $T_{1}/{k}^{\alpha}$ with $\alpha \geq 1$. Indeed, by substitution, $q_{c} = \langle k \rangle / [T_{1}(\langle {k}^{2-\alpha} \rangle - \langle {k}^{1-\alpha} \rangle)]$, that is larger than $1$ because $\langle k \rangle >\langle {k}^{2-\alpha} \rangle \geq \langle {k}^{1-\alpha} \rangle$, and $T_{1} <1$ (assuming $\langle k \rangle \geq 1$). This is actually an equivalent result of that presented in Ref.~\cite{newman1}.

Since $\alpha = 0$ corresponds to standard percolation, we are interested in the intermediate case $0 < \alpha < 1$. 
In scale-free networks the percolation threshold is computed by series summation,
\begin{equation}
q_{c} = \frac{\zeta(\gamma -1)}{\zeta(\gamma+\alpha-2)-\zeta(\gamma+\alpha-1)}~,  
\label{q_alp}
\end{equation}
that has been plotted in Fig.~\ref{figura5}-A as a function of $\gamma$ for different values of $\alpha$ between $0$ and $1$. 
The curves give evidence to the fact that a transition probability moderately decreasing with the degree can induce power-law graphs with $2 < \gamma <3$ to present a finite percolation threshold. However, many real networks (Internet, WWW, etc.) have a power-law degree distribution with exponent lower than $2.5$, for which the giant component persists even for relatively large $\alpha$ (at least in the approximation of infinite systems). 
Nevertheless, the resilience is reduced in finite systems or in presence of a cut-off on the degree as shown by the results of simulations reported in Fig.~\ref{figura5}-B.
The figure shows the behaviour of the threshold value $f_{c}$ as a function of the parameter $\alpha$ for two different exponents $\gamma = 2.3, 2.6$. The lines are the theoretical predictions for (infinite) systems with cut-off $\kappa = 10^2$, the points are numerical results on finite systems with $N=10^5$ and same cut-off.
As for the other results of simulations on power-law graphs that we have presented, the threshold values $f_{c}$ for finite sizes are always larger (i.e. $q_{c}$ is smaller) compared to the theoretical predictions for infinite graphs with the same cut-off on the degree.
This seems to be a characteristic of power-law graphs and would require a separate accurate analysis for increasing sizes $N$ of the graphs.

~\\

This collection of examples is far from being complete, and it should rather represent a way to better understand potential applications of calculations and formulae presented in Section~\ref{mainsec}. In particular, among other possible expressions for the transition probability, we would like to mention non-monotonous or peaked degree-dependent transition probabilities, with  an optimum of transmissibility for (not necessarily large) characteristic values of the degree. 
For a deeper understanding, let us consider the spreading of a virus on the Internet:
highly connected nodes have large exchanges of data with their neighbours, for that they should be potentially the nodes with highest transmission capability. On the other hand, there is a common awareness of their importance, thus they are better protected and controlled, leading to a considerable reduction in the effective transition probability for the  spreading of viruses from these nodes to their neighbours.
Very low-degree nodes are less protected but also less exposed to the transmission (their data exchanges are limited). 
A similar scenario suggests that also non-monotonous transition probabilities are actually interesting in the study of dynamical processes on networks.

\subsubsection{Uniform and random distributed transition probabilities}
\label{sectAv}

In Section~\ref{mainsec}, we have shown that the generalized Molloy-Reed criterion reduces to the results obtained with the intuitive arguments of Section~\ref{naivesec} through successive assumptions. Indeed, as stressed in Ref.~\cite{newman1}, uniform transition probability is far from being an unphysical condition: indeed, in uncorrelated random graphs, $\{T_{ij}\}$ are well approximated by i.i.d. random variables and, in principle, the probability to flow along an edge (bringing physical quantities as information, energy, diseases, etc.) is the average value $\langle T \rangle = T$ over the $\{T_{ij}\}$ distribution.
Hence, we expect that, in random graphs, the average behaviour of site percolation with non-optimal transmission on the edges (with random $T_{ij}$) is well predicted by Eq.~\ref{eq_1}, where $T=\langle T \rangle$ is the average edge transition probability.

\begin{figure} 
\centerline{
\includegraphics*[angle=-90,width=0.5\textwidth]{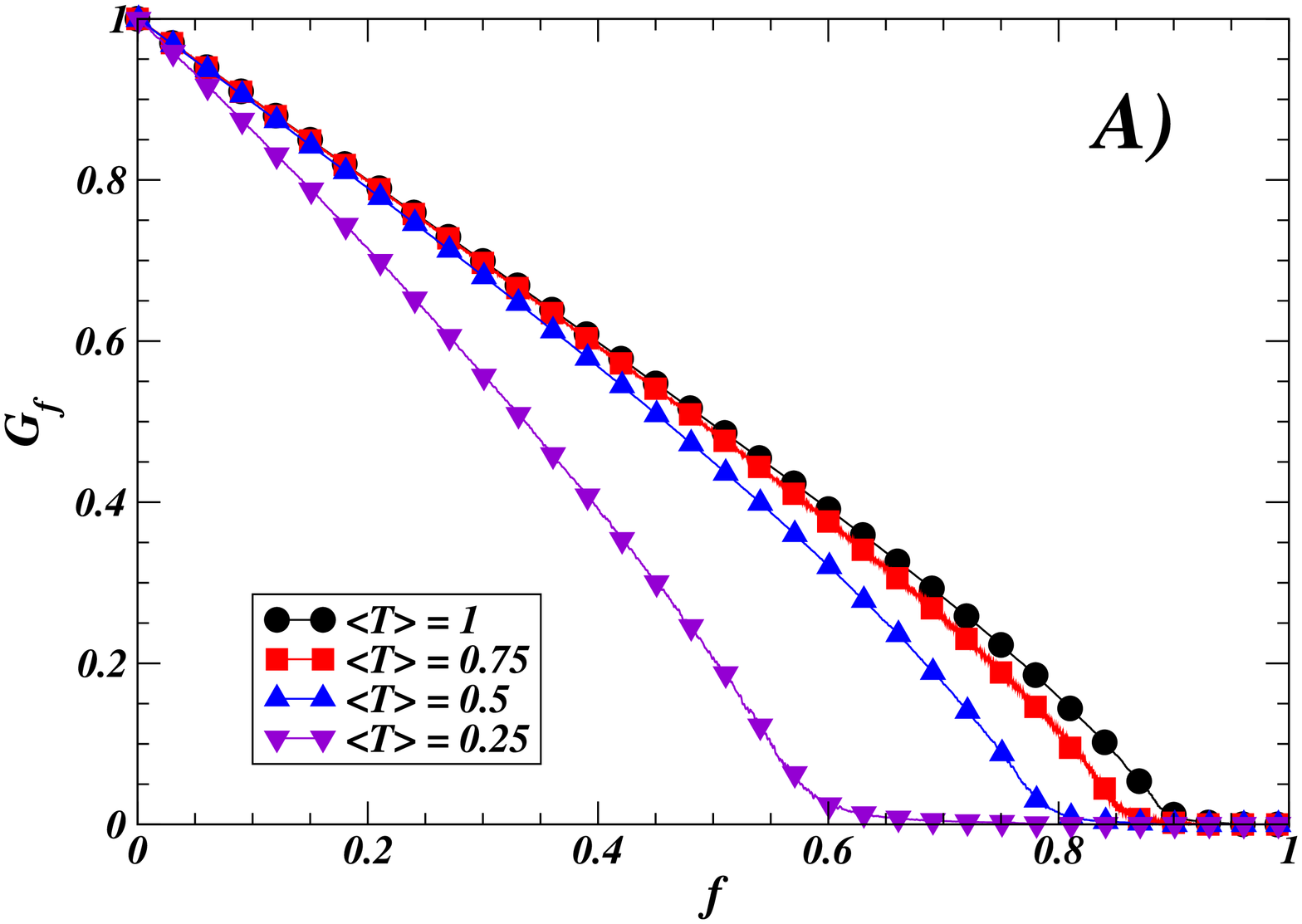}~~
\includegraphics*[angle=-90,width=0.5\textwidth]{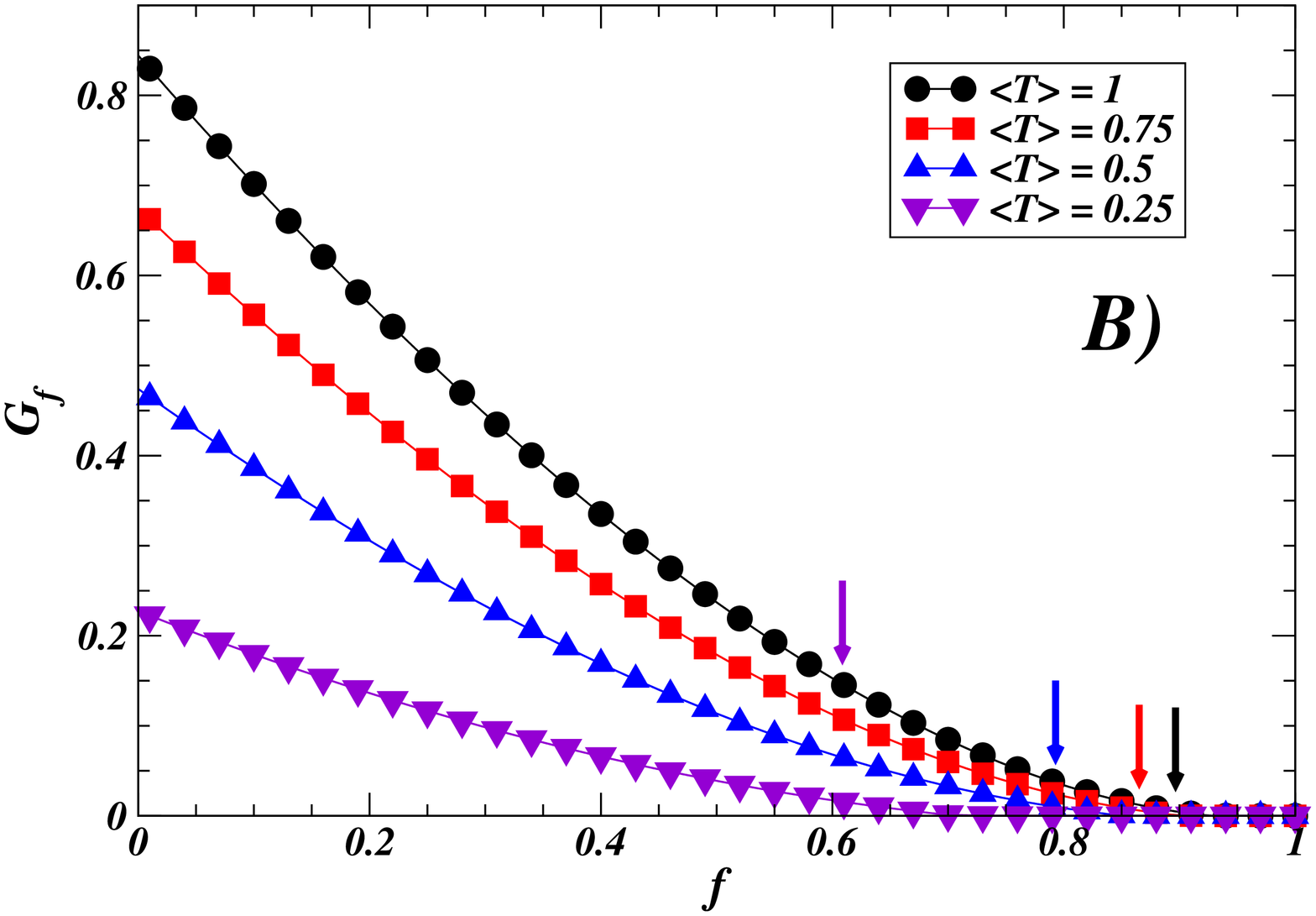}
}
\caption{Size of the giant component $\mathcal{G}_{f}$ as function of the fraction $f$ of removed vertices for $\langle T \rangle = 0.25, 0.5, 0.75, 1$ for an Erd\"os-R\'enyi random graph (A) with $N=10^5$ vertices and $\langle k \rangle = 10$ and a power-law random graphs (B) with same size, exponent $\gamma = 2.3$ and cut-off $\kappa = 10^2$. All curves have been averaged over $100$ realizations. The predicted values of $f_{c} = 0.6, 0.8, 0.867, 0.9$ are well verified in the Erd\"os-R\'enyi graph by the values of $f$ at which $\mathcal{G}_{f} \sim \mathcal{O}(\log N)$. The power-law graph gives slightly worse results, but still in agreement with the theoretical values $f_{c} = 0.613, 0.80, 0.871, 0.9$ (that are indicated by arrows on the curves).\\}
\label{unif}
\end{figure}

This statement has been checked by simulating percolation process on an Erd\"os-R\'enyi random graph with $\langle k \rangle =10$ and on a power-law random graph with exponent $\gamma = 2.3$. Both graphs have $N=10^5$ nodes and the transition probabilities are assigned randomly between $0$ and $1$. Consistently with the data presented in the previous sections, Fig.~\ref{unif}-A displays the behaviour of the network's robustness under random failures. The critical fraction of removed nodes is related to the site percolation's critical occupation probability by $f_{c} = 1- q_{c}$.
Panel A reports the size of the giant component of the Erd\"os-R\'enyi graph as function of the fraction $f$ of removed vertices for $\langle T \rangle = 0.25, 0.5, 0.75, 1$; the predicted values $f_{c} = 0.6, 0.8, 0.867, 0.9$  are well reproduced by simulations. The same measures for a power-law graph are shown in Fig.~\ref{unif}-B. The curves in the simulations are in good agreement with the theoretical values of the percolation threshold $f_{c} \simeq 0.613$ (for $\langle T \rangle = 0.25$), $0.80$ (for $\langle T \rangle = 0.5$), $0.871$ (for $\langle T \rangle = 0.75$), $0.9$ (for $\langle T \rangle = 1.0$). 

%
%
%
 
\subsection{Networks robustness to intentional degree-dependent attacks}

As a final remark on what we have called ``functional robustness'', we should note that other damage strategies, e.g. intentional attacks by degree-driven node removal, can be implemented also in the context of inhomogeneous percolation.
The study of intentional damage to complex networks has been the topic of many research efforts \cite{callaway,cohen3,dorogo1,gallos}, leading to striking analytical and numerical results on the vulnerability of networks.
It has been shown  that scale-free networks are very resilient to random failures, but extremely vulnerable if the attack is brought following a targeted strategy.
In particular, Cohen et. al. \cite{cohen3} have studied analytically the effects of the removal of the most connected vertices, showing that scale-free networks are fragile to degree-driven intentional damage.
When transition probabilities are present, edges are more refractory to transmit physical flows, therefore
scale-free networks fragility should be even more emphasized. 
In order to show this behaviour, we apply to the generalized Molloy-Reed criterion exposed in Eq.~\ref{eq_1k} the semi-numerical method proposed in Ref.~\cite{dorogo1}. We restrict the analysis to two simple cases: intentional attacks to the hubs, and a systematic variation of the attack strategy following a law of preferential selection of the nodes. 

\subsubsection{Attacking the hubs}

Attacks targeted to hit the hubs in networks where the spread of the damage is influenced by edge transition probabilities can, in principle, result in a different threshold behaviour. As a first example, we consider the case in which $T_{ij} = T_{k_{i}} = T_{k}$.
The intentional damage removes vertices with $k > \kappa(f)$, where the cut-off $\kappa(f)$ depends implicitely on the fraction of removed nodes 
\begin{equation}
f = 1 - \sum_{k=1}^{\kappa}p_{k}~;
\label{int_0}
\end{equation}
the condition for the percolation threshold becomes
\begin{equation}
\sum_{k=1}^{\kappa(f)} k (k-1) p_{k} T_{k} = \sum_{k=1}^{\infty} k p_{k}~,
\label{int_1}
\end{equation}

\begin{figure} 
\centerline{
\includegraphics*[angle=-90,width=0.5\textwidth]{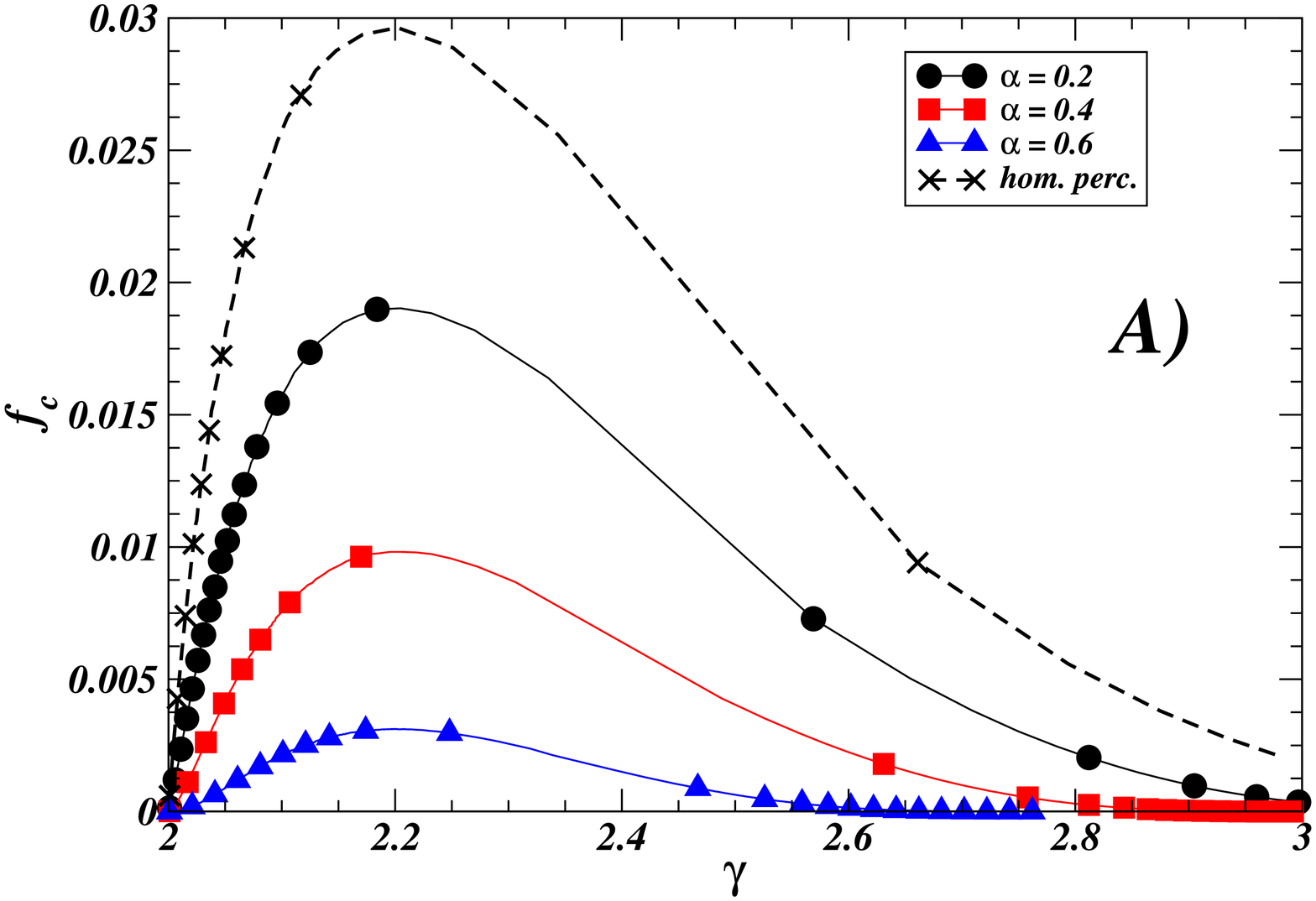}~~
\includegraphics*[angle=-90,width=0.5\textwidth]{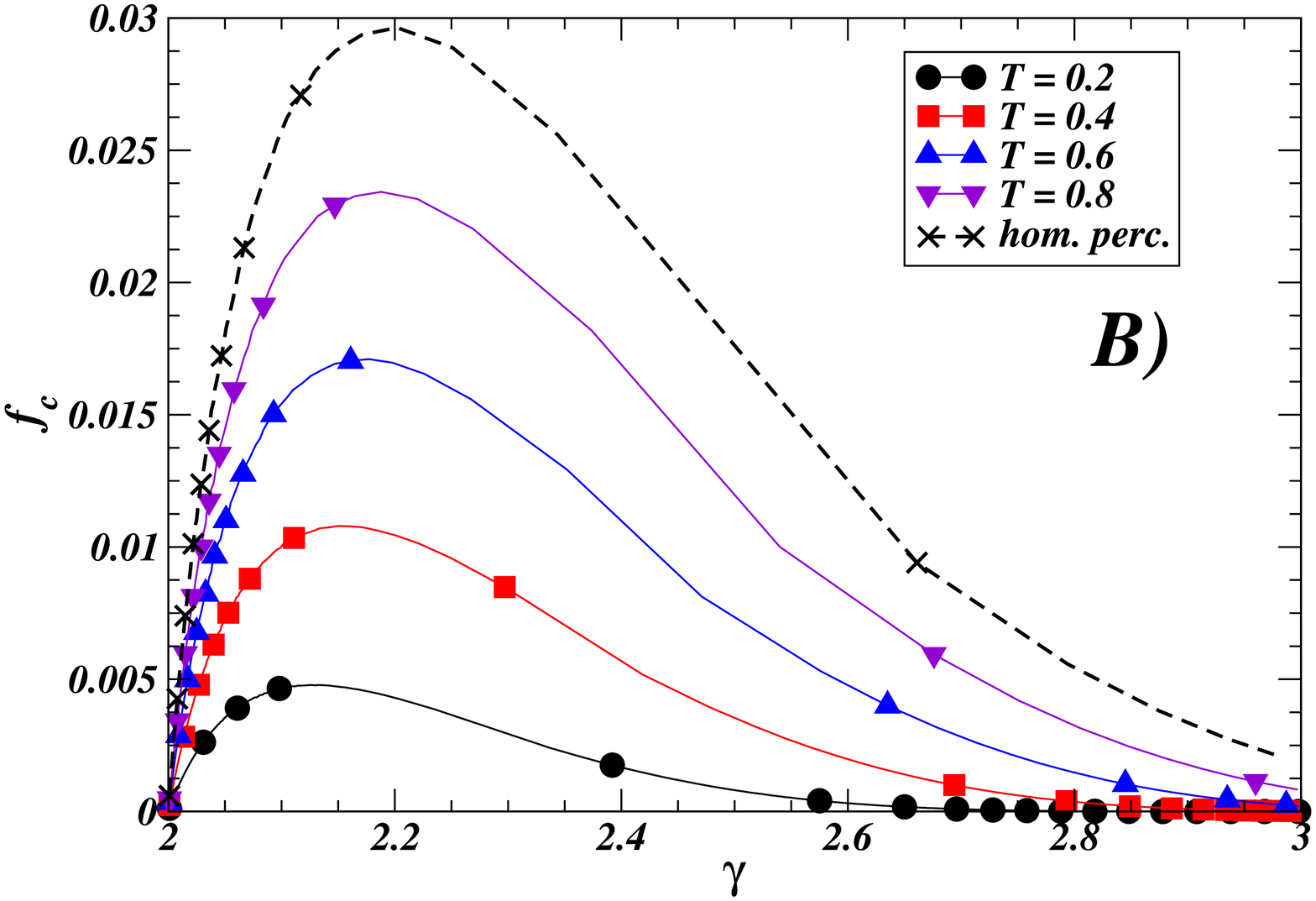}
}
\caption{Fraction $f_{c}$ of removed nodes required to destroy the giant component as a function of the exponent $\gamma$ 
for a power-law graph summitted to intentional attack by removing highest connected nodes. Panel A reports (as full lines) the curves $f_{c}(\gamma)$ when transition probability has the form $T_{k} \sim {k}^{-\alpha}$, with exponent $\alpha = 0.2$ (circles), $0.4$ (squares), $0.6$ (up triangles). Panel B reports the same set of curves for different values of uniform transition probability $T = 0.2$ (circles), $0.4$ (squares), $0.6$ (up triangles), $0.8$ (down triangles). In both panels the case of perfect transmission that corresponds to the standard homogeneous case is plotted in dashed lines.\\}
\label{intent}
\end{figure}

The presence of $T_{k}$ actually modifies the threshold's value $f_{c}$ obtained in the case of optimal transmission. 
According to different functional expressions of $T_{k}$, Eq.~\ref{int_1} provides different behaviours for the critical fraction $f_{c}$. 
All functions $T_{k}$ that are monotonously increasing with $k$ give results consistent with standard percolation. On the other hand, for $T_{k} \sim k^{-\alpha}$ with $\alpha \geq 1$ percolation does not occur; therefore, we consider $0 < \alpha < 1$, that is a transition probability slowly decreasing with $k$. Eq.~\ref{int_1} becomes
\begin{equation}
\sum_{k=1}^{\kappa} k^{2-\gamma-\alpha} = \zeta(\gamma-1) + \sum_{k=1}^{\kappa} k^{1-\gamma-\alpha}~,
\label{int_3}
\end{equation}
that solved numerically for different values of $\alpha = 0.2, 0.4, 0.6$ gives the curves displayed in Fig.~\ref{intent}-A . Their qualitative behaviour is the same as for constant transmissibility. 

Suppose now that transition probabilities $T_{ij}$ do not depend on the degree, but are distributed randomly, then the effective transmissibility is well-approximated by its average value $\langle T \rangle$ (see Section~\ref{sectAv}). Alternatively, one can suppose a uniform transition probability $T=\langle T \rangle$.
Introducing it into Eq.~\ref{int_1} together with the expression of a power-law degree distribution $p_{k} = {\zeta(\gamma)}^{-1}{k}^{-\gamma}$, we find the relation
\begin{equation}
\sum_{k=1}^{\kappa} k^{2-\gamma} = \frac{\zeta(\gamma-1)}{\langle T \rangle} + \sum_{k=1}^{\kappa} k^{1-\gamma}~,
\label{int_2}
\end{equation}
that provides $\kappa(\gamma, \langle T \rangle)$. Then, $f_{c}(\gamma, \langle T \rangle)$ follows from Eq.~\ref{int_0}. 
The equation can be solved numerically, leading to a family of curves for  ${\{f_{c}(\gamma)\}}_{\langle T\rangle}$ that are displayed in Fig.~\ref{intent}-B. The computation is an extension of previous results \cite{dorogo1}, to which our curves tend for $\langle T \rangle \rightarrow 1$. 
%
%
%
It is interesting to note that, in both the cases, decreasing $\langle T \rangle$ implies
a decrease in $f_{c}$, causing the network to be more and more fragile to intentional degree-driven attacks.
     
According to our computations, scale-free networks seem to be ``pathologically'' fragile with respect to degree-dependent removing of nodes, and transition probabilities actually play only a marginal role in determining network's vulnerability under intentional degree-driven attack strategies.

\subsubsection{Systematic variation of attack strategies}
 
Another type of resilience analysis is provided by a numerical study by Gallos et al.~\cite{gallos} on the tolerance of scale-free networks to degree-driven attacks with variable strategies from friendly to intentional attacks. 
A friendly attack consists of the removal of preferentially low-degree nodes, while intentionally attacks are aimed to hit the hubs. The random failure is the intermediate case between these two extremes. 
In Ref.~\cite{gallos}, the systematic variation of attack strategy is encoded into a probability $W(k)$ of choosing a node to be destroyed with a certain fixed probability, that is modeled as a preferential rule
\begin{equation}
W(k) = \frac{k^{\beta}}{\sum_{k=1}^{\kappa} p_{k} {k}^{\beta}}~, 
\label{wk}
\end{equation}
with $-\infty < \beta < +\infty$. The authors shows numerically that for $\beta >0$ the effects of the attacks are progressively more destructive as $\beta$ is increased, while for $\beta<0$ the fraction of removed nodes $f$ necessary to destroy the network is very close to $1$. 

In our context, preferential attack strategies can be associated to a bond percolation process with power-law occupation (or traversing) probability $q_{k}$. In terms of percolation, indeed, the fraction $f$ of removed nodes is written 
\begin{equation}
f=1-\sum_{k=1}^{\kappa} p_{k}q_{k}~,
\label{fW}
\end{equation}
with $q_{k} \propto 1-W(k)$; then the cut-off $\kappa(f)$ is a function of $f$ as in Eq.~\ref{int_0}. 
The correct form of the generalized Molloy-Reed criterion comes directly from Eq.~\ref{eq_1k} with $T_{k}=1$
$\forall k$, and states that a giant component exists when
\begin{equation}
\sum_{k=1}^{\kappa(f)} k [(k-1)q_{k} - 1] p_{k} \geq 0~.  
\label{mrcgallos}
\end{equation}
Requiring the equality, it is possible to compute numerically the threshold value $f_{c} =f_{c}(\gamma, \beta)$ of removed nodes. 
Very recently, Gallos et al. (\cite{gallos2}) have obtained an approximated expression for the percolation condition in the case the attacks are sequentially brought. They have evaluated the probability $\rho(k) = {(1-W(k))}^{d}$ that a node of degree $k$ is still present after $d$ node removals. However, if we assume that the attacks are simultaneously launched, the only important quantity is the fraction $f$ of removed nodes, while the survival probability is $\rho(k) = (1-W(k)) = q_{k}$ ($d=1$). In this limit the previously exposed approach is completely equivalent to that in Ref.~\cite{gallos2}. 
An interesting improvement of this method should be the study of networks resilience in the case of inhomogeneous edges ($T_{k} \neq 1$).

~\\

\section{CONCLUSIONS}
\label{concsec}

Percolation theory provides powerful methods to study statistical properties of physical systems and, recently, it has been exploited to uncover structural features of real complex networks such as the Internet. 
Nevertheless, standard homogeneous percolation fails in pinpointing the interplay between the topology and the properties of the flows and other dynamical phenomena taking place on it. Typical examples are spreading phenomena: frequently modeled by (homogeneous) percolation, the real processes possess features that cannot be encoded in its formalism because spreading rates are non homogeneous and depend on nodes and links characteristics.
To model more realistic processes, in which the flow of information, energy, or any other physical quantity depend on the transmission ability of nodes and edges, we have introduced inhomogeneous edge transition probability and node occupation probability and studied the corresponding joint site-bond percolation process. The influence of inhomogeneity is reflected not only in the structural properties but also in the network's functionality and efficiency.

Inhomogeneous percolation has been analytically studied on random networks with generating functions techniques, obtaining explicit expressions for the site and bond percolation thresholds as a function of both the topology and the transition probability.
The main result is the generalization of the well-known Molloy-Reed criterion for the existence of a giant component (i.e. of percolation) for a Markovian correlated random graph with arbitrary degree distribution. The analytical study considers also the very general case of multi-state nodes, that seems to be interesting for social networks, where the existence of community structures is well-established. Simpler situations regarding correlated and uncorrelated random graphs composed by identical nodes have been also studied. In particular, explicit expressions for the critical threshold of the site percolation with inhomogeneous edges have been computed in the case of uncorrelated random graphs.

In addition, percolation properties have been used to study the vulnerability of uncorrelated random graphs to random failures revealing that their functional robustness can be seriously reduced if the disparity of edge transition probabilities are taken into account in computing the effective giant component.
Analytical results have been checked by numerically simulating percolation on homogeneous and heterogeneous uncorrelated random networks. 

Many real weighted networks such as infrastructure or technological networks, present a relation between the weights and the actual flow of physical quantities (information, energy, goods, etc.) carried by the links.
For this reason, they seem to represent a reasonable application's field for inhomogeneous percolation. Furthermore, a better knowledge of real data about Internet's traffic, diseases spreading and other flows on networks could provide major insights on what type of transition probability functions are actually relevant. 

In conclusion, beside the purely theoretical interest for a generalization of percolation on random graphs and its analytical solution, this paper provides a method to incorporate into the  structural properties some aspects of the functional description of networks.
Therefore, the implications of this new approach go beyond the details of the present study, and interest directly the recent and very stimulating issue of the interaction between network's topology and dynamical phenomena occurring on it.

\section*{Acknowledgments}

The author is particularly grateful to A. Barrat for illuminating discussions and the careful reading of the manuscript.
The work has been partially supported by the European Commission - Fet Open Project COSIN IST-2001-33555 and contract 001907 (DELIS).

\appendix

\section{Generating functions in percolation problems}
\label{append1}

In order to better specify the notations used in the main text, we recall some basic notions about generating functions,
considering only the case of infinite graphs without isolated vertices. 
The generating function for the degree distribution $p_k$ of a randomly chosen vertex is
\begin{equation}
G_{0}(x)=\sum_{k=1}^{\infty}p_{k}x^{k}~,
\end{equation}
with $G_{0}(1)=1$, and $\langle k \rangle =\sum_{k}k p_{k} = G_{0}'(1)$. 
Similarly, the generating function for the probability that a randomly chosen edge leads to a vertex of given degree is
\begin{equation}
\frac{\sum_{k}k p_{k} x^{k-1}}{\sum_{k}k p_{k}} = G_{1}(x)= \frac{G_{0}'(x)}{G_{0}'(1)}~.
\end{equation}
A useful property is that the probability distribution and its moments can be computed by simple derivative of the corresponding generating function. 

If we call $q_{k}$ the probability that a vertex of degree $k$ is occupied (or node traversing probability if regarded as a spreading phenomena), the probability that, choosing randomly a vertex, we pick up an occupied vertex of degree $k$ is the product of the probabilities of two independent events, i.e. $p_{k}q_{k}$. Repeating the same operation with the edges, we need the probability that the randomly chosen edge is attached to an occupied vertex of degree $k$. This event happens with probability $k p_{k} q_{k} /\langle k \rangle$.  
Hence, we define the generating functions for both these probabilities that are very important in the site percolation,
\begin{subequations}
\begin{align}
 F_{0}(x; \{q\})& =\sum_{k=1}^{\infty}p_{k} q_{k} x^{k}~,\label{f0}\\
 F_{1}(x; \{q\})& =\frac{\sum_{k=1}^{\infty} k p_{k} q_{k} x^{k-1}}{\sum_{k}k p_{k}} = \frac{F_{0}'(x)}{\langle k\rangle}~.\label{f1}
 \end{align}
\end{subequations}
The function $F_{0}(x; \{q\})$ is the generating function of the probability that a vertex of a given degree exists and is occupied, while $F_{1}(x; \{q\})$ is the generating function for the probability of reaching a vertex of a given degree starting by a randomly chosen edge and that it is occupied.

The solution of the site percolation problem is the set of values $\{q_k\}$ for which an infinite cluster (giant component) exists. The same analysis holds if we collect the nodes on the basis of other properties, but in the following only the degree-dependent formalism will be developed. 

In order to compute the probability that a randomly chosen vertex belongs to the giant component, we start by computing the probability $P_{s}$ that a randomly chosen vertex belongs to a connected cluster of a certain size $s$. 
The use of generating functions allows to do it simultaneously for all the possible sizes. Then, the mean cluster size is obtained as the first derivative of the generating function of $P_{s}$. 
Finally, the condition for the divergence of the mean cluster size gives the condition for the existence of the giant component as a function of the parameters of the system, that are the degree distribution and the node occupation probability.

Firstly, we consider the probability $P_{s}$ that a randomly chosen vertex in the network belongs to a connected cluster of a certain size $s$. We call $H_{0} (x; \{q\})$ its generating function,
\begin{equation}
H_{0}(x; \{q\}) = \sum_{s=0}^{\infty} P_{s} x^{s}~,
\end{equation}
in which we have conventionally grouped in the term for $s=0$ the probability $1-\sum_{k}q_{k}p_{k}$ that a vertex is not occupied.
Similarly, let $\hat{P_{s}}$ be the probability that a randomly chosen edge leads to a cluster of a given size $s$, and $H_{1}(x; \{q\})$  its generating function. 
  
\begin{figure} 
\centerline{
\begin{tabular}{|c|}\hline \\ \includegraphics*[angle=-90,width=0.45\textwidth]{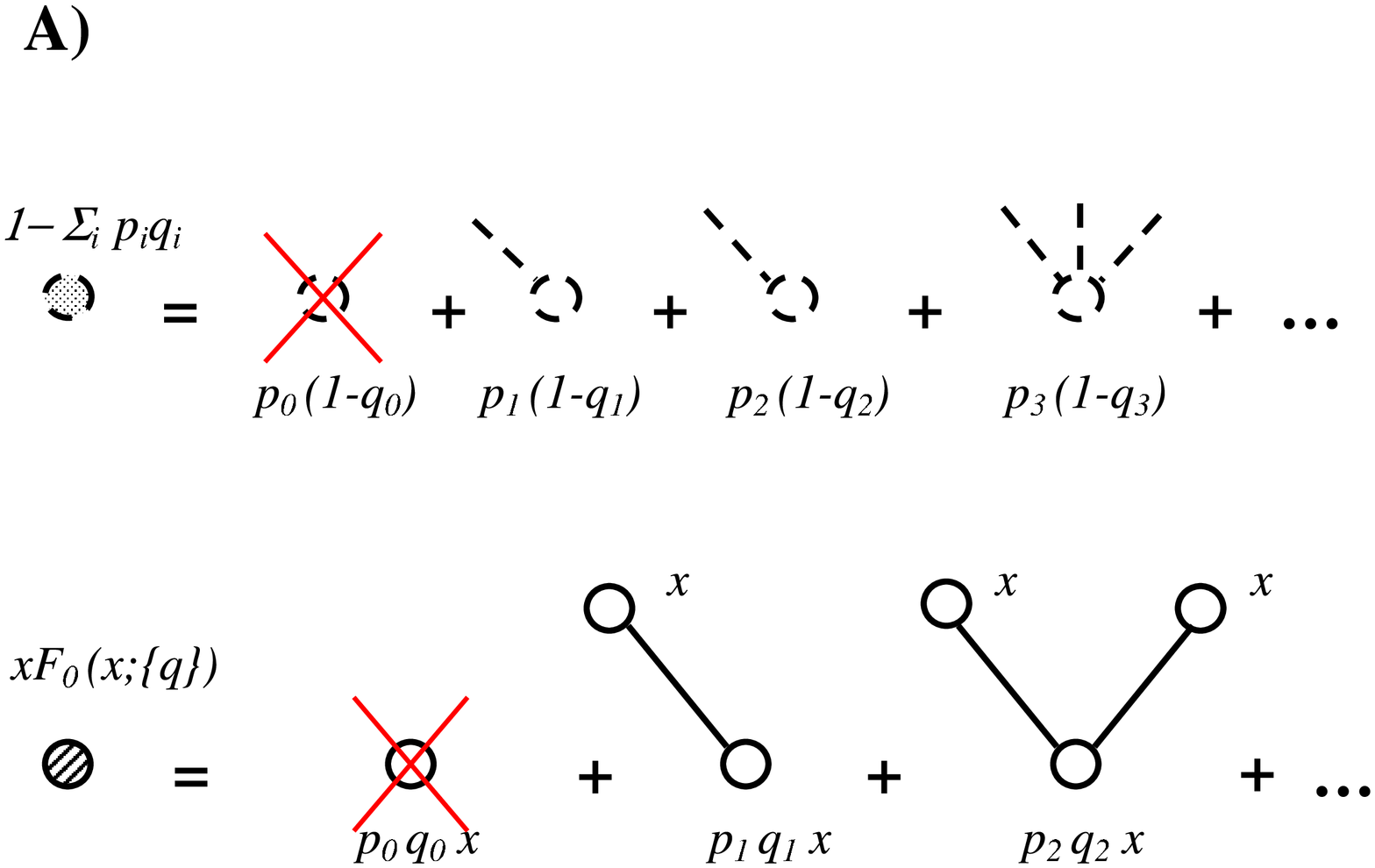}\\ \hline\end{tabular}~~
\begin{tabular}{|c|}\hline \\ \includegraphics*[angle=-90,width=0.45\textwidth]{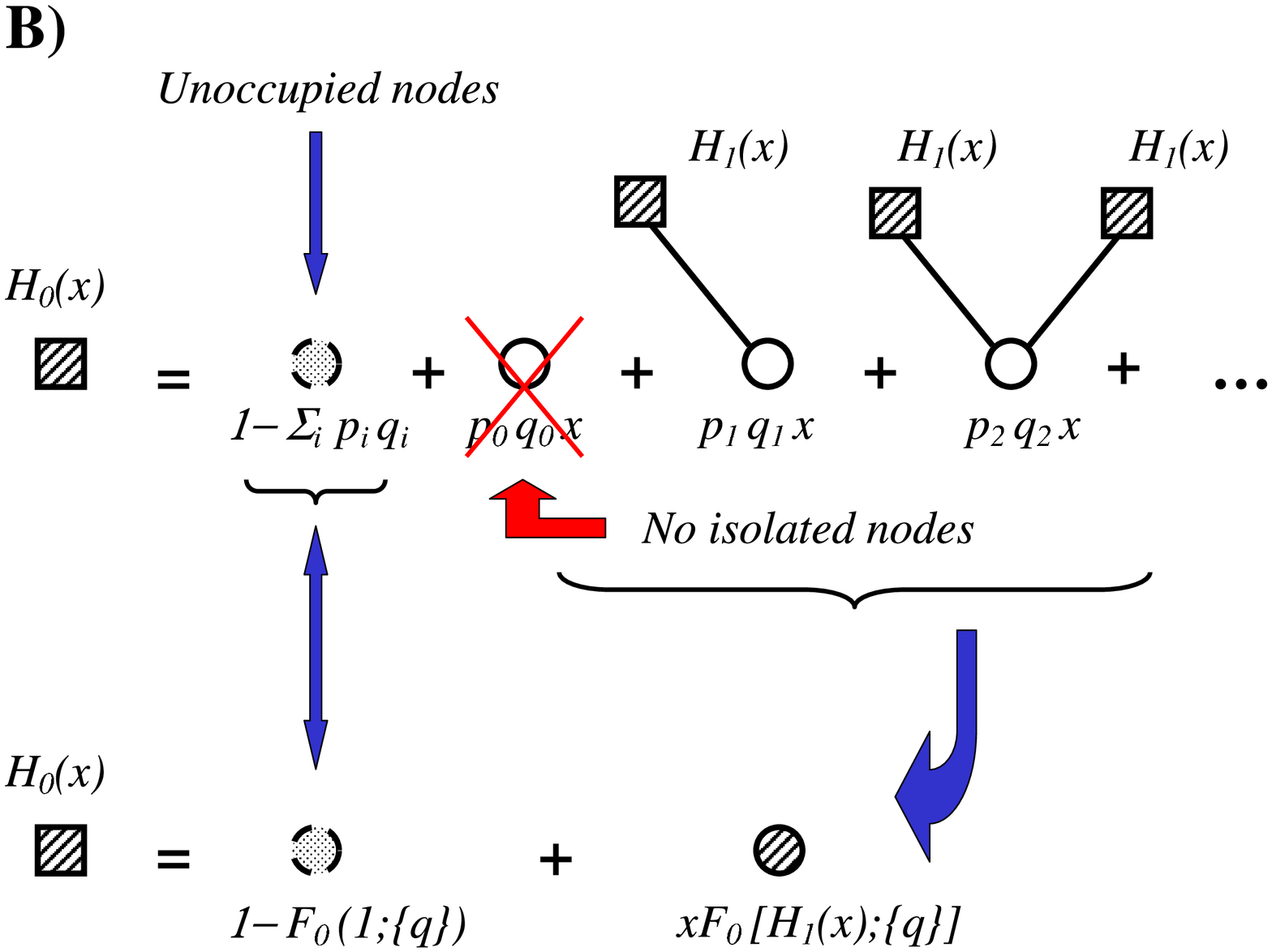}\\ \hline\end{tabular}
}
\caption{A) A full dotted bullet with dashed contour line corresponds to the probability that a vertex is unoccupied. This is given by a particular series of diagrams, in which we sum the contributions of unoccupied vertices of all possible degrees.  
A striped bullet with full contour line represents the generating function $F_{0}(x;\{q\})$, whose diagrammatical expansion
contains all possible combinations of occupied vertices reachable by an occupied vertex with a certain degree. The $x$ accounts for the occupation of a vertex. 
The contributions of the isolated vertices ($p_{0}q_{0}$) have been deleted in agreement with our convention of considering only graphs with non isolated vertices.  
B) Diagrammatical representation of the generating function $H_{0}(x)$. The first terms of the infinite series correspond to the summation of the Eqs.~\ref{series} as presented in Eq.~\ref{lungaH0}. These contributions can be expressed in a compact form using Eq.~\ref{h00}, that contains the  generating functions $F_{0}(x;\{q\})$ and $H_{1}(x)$.\\}
\label{diag1}
\end{figure}

Since we choose the starting vertex at random, each possible degree $k$ gives a different contribution to each possible cluster size probability $P_{s}$, meaning that each term $P_{s} x^{s}$ is itself given by an infinite sum of terms labeled by the degree $k$. 
For instance, the first terms of $H_{0}(x; \{q\})$ are:
\begin{subequations}\label{series}
\begin{align}
 P_{0}~~& = 1 - \sum_{k} p_{k} q_{k}~,\label{ser1}\\
 P_{1}x~& = \sum_{l=1}^{\infty} {p_{l} q_{l} x ~\hat{P_{0}}}^{l}~,\label{ser2}\\
 P_{2}{x}^{2}& = \sum_{l=1}^{\infty} {l p_{l} q_{l} x ~(\hat{P_{1}} x)}~{\hat{P_{0}}}^{l-1}~,\label{ser3}\\
 P_{3}{x}^{3}& = \sum_{l=1}^{\infty} {l p_{l} q_{l} x ~(\hat{P_{2}} {x}^{2})~\hat{P_{0}}^{l-1}} + \sum_{l\geq 2} p_{l} q_{l} x ~{\left(\begin{array}{c} l\\ 2\end{array}\right)}~{(\hat{P_{1}} x)}^{2}~\hat{P_{0}}^{l-2}~,\label{ser4}\\
 P_{4}{x}^{4}& = \sum_{l=1}^{\infty} l p_{l} q_{l} x ~(\hat{P_{3}} {x}^{3})~{\hat{P_{0}}}^{l-1} + \sum_{l\geq 2} p_{l} q_{l} x ~2 {\left(\begin{array}{c} l \\ 2 \end{array} \right)}~{(\hat{P_{2}} x^2)}~(\hat{P_{1}} x)~{\hat{P_{0}}}^{l-2} + \sum_{l\geq 3} p_{l} q_{l} x ~{\left(\begin{array}{c} l \\ 3 \end{array} \right)}~{(\hat{P_{1}} x)}^3~{\hat{P_{0}}}^{l-3}~,\label{ser5}\\
\nonumber ~& \vdots
\end{align}
\end{subequations}
Now, summing these terms and grouping similar contributions we get
\begin{equation}\label{lungaH0}
\begin{split} H_{0}(x; \{q\})& = 1 - \sum_{k} p_{k} q_{k} + p_{1} q_{1} x ~ \left[ \hat{P_{0}} + \hat{P_{1}} x + \hat{P_{2}} {x}^{2} + \dots \right] \\
&\quad + p_{2} q_{2} x ~ \left[ {\hat{P_{0}}}^{2} + 2 \hat{P_{0}} ~\hat{P_{1}} x + {(\hat{P_{1}} x)}^{2} + 2 \hat{P_{0}} ~\hat{P_{2}} x^{2} + 2 \hat{P_{1}} x ~ \hat{P_{2}} {x}^{2} + {(\hat{P_{2}} {x}^{2})}^{2} + \dots \right] + \dots \\
&= 1 - \sum_{k} p_{k} q_{k}  + p_{1} q_{1} x \left[ \hat{P_{0}} + \hat{P_{1}} x +  \hat{P_{2}} {x}^{2} + \dots \right] + p_{2} q_{2} x ~{\left[ \hat{P_{0}} + \hat{P_{1}} x +  \hat{P_{2}} {x}^{2} + \dots \right]}^{2} + \dots~\\
&= 1 - \sum_{k} p_{k} q_{k}  + x p_{1} q_{1} H_{1}(x; \{q\}) + x p_{2} q_{2} {\left[ H_{1}(x; \{q\})\right]}^{2} + \dots~.
\end{split}
\end{equation}
Note that no term contains $p_{0}q_{0}$ according with the convention of considering only non isolated nodes.
A compact form for this expression is written using Eq.~\ref{f0},
\begin{eqnarray}
H_{0}(x; \{q\}) = 1-F_{0}(1;\{q\}) + x F_{0}(H_{1}(x; \{q\}); \{q\})~.
\label{h00}
\end{eqnarray}
The structure of Eq.~\ref{h00} can be represented diagrammatically as shown in Fig.~\ref{diag1}, associating the variable $x$ to each ``bare'' vertex and a variable $H(x; \{q\})$ to each ``dressed'' vertex, while the function $F_{0}$  weights the contributions over all possible degrees.  

%
%
%
%
Moreover, the generating function $H_{1}(x; \{q\})$ satisfies a similar self-consistent equation,
\begin{eqnarray}
H_{1}(x; \{q\}) = 1-F_{1}(1;\{q\}) + x F_{1}(H_{1}(x; \{q\}); \{q\})~,
\label{h01}
\end{eqnarray}
that is obtained following completely similar arguments starting from picking up an edge at random. 

The mean cluster size $\langle s \rangle$ is computed directly from Eq.~\ref{h00} as the first derivative of $H_{0}$ with respect to $x$ in $x=1$. Imposing its divergence allows to find the condition for the existence of a giant component, that corresponds to the Molloy-Reed criterion as presented in Ref.~\cite{callaway}. Finally, considering a uniform occupation probability $q_{k}=q$ (or uniform node traversing probability), this criterion gives the expression for the site percolation threshold $q_{c}$. 

\section{Inhomogeneous percolation in graphs with multi-state nodes}
\label{append2}

Consider a Markovian correlated graph with multi-state vertices, i.e. the vertices are divided into $n$ different classes or states, and suppose that each vertex is given an occupation probability depending on its degree and each edge is endowed with a transition probability depending on the degrees and the states of the extremities. 
The fundamental brick for the construction of generating functions in correlated graphs is the rooted edge composed of a starting vertex $i$ and the pending edge $(i, j)$ connecting it to a second vertex $j$, without explicitly considering this second extremity. For this reason we will always average on the degree of the second extremity of the edge.  
Let us consider a vertex $i$ chosen at random, it will be characterized by a class $h$ and by a degree $k_{i}$. In principle, the $k_{i}$ edges departing from that node are connected to $k_{i}$ other nodes belonging to different classes. Actually, only $m_{i}$ of them are really reached by a flow  because of the presence of transition probabilities on the edges (we call such transmitting edges {\em open}). Therefore, they identify a partition of $\{m_{i}^{(1)}, m_{i}^{(2)}, \dots, m_{i}^{(n)}\}$, with $\sum_{l} m_{i}^{(l)} = m_{i} \leq k_{i}$, in which $m_{i}^{(l)}$ is the number of these neighbouring nodes belonging to the class $l$ and linked to $i$ by an open edge.

Suppose that $m_{i}^{(l)}$ of the $k_{i}$ edges emerging from a node of class $h$ and degree $k_{i}$ are successfully connected to nodes of a same class $l$ and (possibly different) degrees $k_{j}$. The average probability that an edge among them allows the flow to pass is $\sum_{k_{j}} T^{(h \rightarrow l)}_{k_{i} k_{j}} p^{(h\rightarrow l)}(k_{j}|k_{i})$, where $p^{(h\rightarrow l)}(k_{j}|k_{i})$ is the degree conditional probability between vertices of states $h$ and $l$ and $T^{(h\rightarrow l)}_{k_{i} k_{j}}$ is the transition probability along an edge from a node of degree $k_{i}$ in the class $h$ to a node of degree $k_{j}$ in the class $l$.
The origin of this term is trivial: the probability to pass along the edge is the product of two independent events, i.e the edge exists and it is open; then, being interested in rooted edges, we have to average over all possible degrees $k_{j}$. 
The probability that there are $m_{i}^{(l)}$ of these edges produces a term ${[\sum_{k_{l}} T^{(h\rightarrow l)}_{k_{i} k_{j}} p^{(h\rightarrow l)}(k_{j}|k_{i})]}^{m_{i}^{(l)}}$. 
Positive events give $n$ contributions of this kind, while the $k_{i}-m_{i}$ negative events contribute to a single term ${\left[ 1- \sum_{l=1}^{n} \sum_{k_{j_{l}}} T^{(h\rightarrow l)}_{k_{i} k_{j_{l}}} p^{(h\rightarrow l)}(k_{j_{l}}|k_{i}) \right]}^{k_{i}-m_{i}}$, that is the probability that $k_{i} - m_{i}$ edges do not admit the flow's passage whichever class they belong to.
Computing the probability of the whole event associated with the {\em partition} $\{m^{(0)}_{i}=k_{i}-m_{i}, m^{(1)}_{i}, m^{(2)}_{i}, \dots, m^{(n)}_{i}\}$ of the neighbours of the node with degree $k_{i}$, we get the multinomial distribution 
\begin{equation}
P^{(h)}(k_{i},\{m^{(l)}_{i}\}) = {k}_{i}! \frac{1}{m_{i}^{(0)}!} {\left[ 1- \sum_{j_{l}=1}^{n} \sum_{k_{j_{l}}} T^{(h\rightarrow l)}_{k_{i} k_{j_{l}}} p^{(h\rightarrow l)}(k_{j_{l}}|k_{i}) \right]}^{m^{(0)}_{i}} \times \prod_{l=1}^{n} \frac{1}{m^{(l)}_{i}!} {\left[ \sum_{k_{j_{l}}} T^{(h\rightarrow l)}_{k_{i} k_{j_{l}}} p^{(h\rightarrow l)}(k_{j_{l}}|k_{i}) \right]}^{m^{(l)}_{i}}. 
\label{multi1}
\end{equation}
A simpler version of this multinomial distribution appears in Ref.~\cite{newman4}.

The following step consists in using the expression of the multinomial distribution to obtain the generating function of the probability that a physical quantity spreading from a vertex of class $h$ successfully flows through $\{m^{(l)}\}$ of its edges that point to vertices in the class $\{l\}$ ($l=1,2,\dots,n$). Summing over all possible values of $k_{i}$ and over all possible partitions of $k_{i}$ in $n+1$ values $\{m^{(l)}\}$, we obtain the generating function 
\begin{equation}
F_{0}^{(h)}(x_{1},x_{2}, \dots, x_{n};\{q, T\}) = \sum_{k_{i}=1}^{\infty} p_{k_{i}}^{(h)} q_{k_{i}}^{(h)} \sum_{\{m^{(l)}_{i}\}} \delta(k_{i}, \sum_{l=0}^{n} m^{(l)}_{i}) P^{(h)}(k_{i}, \{m^{(l)}_{i}\}) \prod_{l=1}^{n} x_{l}^{m^{(l)}_{i}}, 
\label{multi2}
\end{equation}
in which $q_{k_{i}}^{(h)}$ is the occupation (traversing) probability of a vertex belonging to the class $h$ with degree $k_{i}$, $\delta(\cdot,\cdot)$ is a Kronecker's symbol and the $x_{1}, \dots, x_{n}$ variables represent the average contributions of the rooted edges of the different classes.
Introducing Eq.~\ref{multi1} in Eq.~\ref{multi2}, the sum over the partitions $\{m^{(l)}_{i}\}$ corresponds to the extended form of a multinomial term, providing the following expression for the generating function  
\begin{equation}
F_{0}^{(h)}(x_{1},x_{2}, \dots, x_{n};\{q, T\}) = F_{0}^{(h)}(\mathbf{x};\{q, T\}) = \sum_{k_{i}=1}^{\infty} p_{k_{i}}^{(h)} q_{k_{i}}^{(h)} {\left[ 1+ \sum_{l=1}^{n} (x_{l}-1) \sum_{k_{j_{l}}} T^{(h\rightarrow l)}_{k_{i} k_{j_{l}}} p^{(h\rightarrow l)}(k_{j_{l}}|k_{i}) \right]}^{k_{i}}. 
\label{multi3}
\end{equation}

With a completely similar argument, we can compute the generating function $F_{1}^{(h)}(\mathbf{x};\{q,T\})$ of the probability that a randomly chosen edge leads to a vertex of class $h$ from which the
spread toward its neighbours successfully flows through $\{m^{l}\}$ edges pointing to nodes of class $\{l\}$ ($l=1,2,\dots,n$). 
Hence, observing that now the number of emerging edges available to the spreading process reduces to $k_{i}-1$ and that the probability to reach the starting vertex (from an edge pointing to a generic vertex of class $h$) is $\frac{k_{i} p_{k_{i}}^{(h)}}{\sum_{k} k p_{k}^{(h)}}$, the generating function $F_{1}^{(h)}(\mathbf{x};\{q, T\})$ reads
\begin{equation}
F_{1}^{(h)}(x_{1},x_{2}, \dots, x_{n};\{q, T\}) = F_{1}^{(h)}(\mathbf{x};\{q, T\}) = \sum_{k_{i}=1}^{\infty} \frac{k_{i} p_{k_{i}}^{(h)}}{\sum_{k} k p_{k}^{(h)}} q_{k_{i}}^{(h)} {\left[ 1+ \sum_{l=1}^{n} (x_{l}-1) \sum_{k_{j_{l}}} T^{(h\rightarrow l)}_{k_{i} k_{j_{l}}} p^{(h\rightarrow l)}(k_{j_{l}}|k_{i}) \right]}^{k_{i}-1}. 
\label{multi4}
\end{equation}

As recalled in the previous subsection, the two generating functions are useful in the computation of a system of self-consistent equations (similar to those in Eqs.~\ref{h00}-\ref{h01}) from which the expression of the average cluster size $\langle s \rangle$ should be derived. The main difference concerns the form of the generating functions $F_{0}^{(h)}(\mathbf{x};\{q, T\})$ and $F_{1}^{(h)}(\mathbf{x};\{q, T\})$, that are partitioned in classes (of nodes in different states) and contain the contributions of the transition probabilities. Firstly, we consider the probability $P_{s}^{(h)}$ that a randomly chosen edge leads to a vertex of class $h$ belonging to a connected component of a given size $s$. Its generating function $H_{1}^{(h)}(x;\{q, T\}) = \sum_{s} \hat{P}_{s}^{(h)} x^{s}$ satisfies the self-consistent equation
\begin{equation}
H_{1}^{(h)}(x; \{q, T\}) = 1-F_{1}^{(h)}(\mathbf{1};\{q, T\}) + x~ F_{1}^{(h)}[H_{1}^{(1)}(x;\{q, T\}), \dots, H_{1}^{(n)}(x;\{q, T\});\{q, T\}]~,
\label{multi5}
\end{equation}
where the presence of $H_{1}^{(h)}(x;\{q, T\})$ for all $h=1, \dots, n$ on the r.h.s. means that the constraint on the value of $h$ is required only on the starting node, not on the others reachable from it. Moreover, $x$ refers to the cluster distribution and does not need any label. 

The first term in the r.h.s. of Eq.~\ref{multi5} is due to the probability that the node of class $h$ to which a chosen edge leads is not occupied, therefore it should not depend on the transmissibility of any outgoing edge. As required, the term $1-F_{1}^{(h)}(\mathbf{1};\{q, T\})$ computed in $x=1$ does not depend on $\{T\}$. This corresponds exactly to the term $\hat{P}_{0}^{(h)}$ in the cluster expansion. The second term of Eq.~\ref{multi5} refers to the contribution of an occupied vertex. Let us suppose that its degree is $k_{i}$ and consider one of its outgoing edges leading to a vertex in the class $l$: its contribution is given by the probability $1-\sum_{k_{j_{l}}} T^{(h\rightarrow l)}_{k_{i} k_{j_{l}}} p^{(h\rightarrow l)}(k_{j_{l}}|k_{i})$ that the flow does not reach the second extremity of the edge, and by the probability that it passes ($\sum_{k_{j}} T^{(h \rightarrow l)}_{k_{i} k_{j}} p^{(h\rightarrow l)}(k_{j}|k_{i})$). The latter has to be multiplied by the vertex function associated to the probability that this second vertex (of class $l$) belongs to a cluster of a given size. 
This probability is generated by the function $H^{(l)}_{1}(x; \{q, T\})$ that is the correct vertex term for this contribution.
Then, all these quantities have to be averaged over the set of degrees $k_{i}$, with weights $p^{(h)}_{k_{i}}$ and $q^{(h)}_{k_{i}}$, easily recovering the second term on the r.h.s.. Since the spirit of the derivation is completely the same, we refer again to Fig.~\ref{diag1} for a diagrammatical representation of the Eq.~\ref{multi5} (details are different). 
 
The other equation, for the generating function $H_{0}^{(h)}(x; \{q, T\})$ of the probability that a randomly chosen vertex of class $h$ belongs to a cluster of fixed size $s$, reads
\begin{equation}
H_{0}^{(h)}(x; \{q, T\}) = 1-F_{0}^{(h)}(\mathbf{1}; \{q, T\}) + x~ F_{0}^{(h)}[H_{1}^{(1)}(x; \{q, T\}), \dots, H_{1}^{(n)}(x; \{q,T\}); \{q, T\}]~,
\label{multi6}
\end{equation}
Note that, by definition, both $H_{0}^{(h)}(x; \{q, T\})$ and $H_{1}^{(h)}(x; \{q, T\})$ are $1$ in $x=1$ for all $h$.

Now, taking the derivative of Eq.~\ref{multi6} with respect to $x$ in $x=1$, we obtain the average number of vertices reachable starting from a vertex in the class $h$,
\begin{equation}\label{multi7}
\begin{split} \langle s_{h} \rangle & = \frac{d H_{0}^{(h)}}{d x}{\bigg \vert}_{x=1}\\  
&= F_{0}^{(h)}(\mathbf{H}_{1}(1)=\mathbf{1}; \{q, T\}) + \sum_{l} \frac{\partial F_{0}^{(h)}}{\partial x_{l}}{\bigg \vert}_{\mathbf{x}=\mathbf{1}} {H_{1}^{(l)}}'(1; \{q, T\})~.
\end{split}
\end{equation}
The second term in the r.h.s. contains linear contributions from other classes of vertices, therefore Eq.~\ref{multi7} can be written in a matrix form (in the $n\times n$ product space generated by pairs of classes) as
\begin{equation}
\langle \mathbf{s} \rangle = \nabla_{x} \mathbf{H}_{0}(x; \{q, T\}) {\bigg \vert}_{x=1} = \mathbf{F}_{0}[\mathbf{1}; \{q, T\}] + \nabla_{x} \mathbf{F}_{0}[\mathbf{x}; \{q, T\}]{\bigg \vert}_{\mathbf{x}=\mathbf{1}} \cdot \nabla_{x} \mathbf{H}_{1}(x;\{q, T\}){\bigg \vert}_{x=1}~,
\label{multi8}
\end{equation}  
with $\mathbf{s} = (s_{1}, s_{2}, \dots, s_{n})$.
Taking the derivative of Eq.~\ref{multi5} with respect to $x$ in $x=1$, we obtain an implicit expression for ${H_{1}^{(h)}}'(1; \{q, T\})$ 
\begin{equation}
{H_{1}^{(h)}}'(1; \{q, T\}) = F_{1}^{(h)}[\mathbf{1}; \{q, T\}] + \sum_{l} \frac{\partial}{\partial x_{l}} F_{1}^{(h)}[1; \{q, T\}] {H_{1}^{(l)}}'(1; \{q, T\})~, 
\label{multi9}
\end{equation}
and putting together the contributions in a matrix formulation, we get
\begin{equation}
{\nabla_{x} \mathbf{H}_{1}(x; \{q, T\})}{\bigg \vert}_{x=1}  = \mathbf{F}_{1}[\mathbf{1}; \{q, T\}] + \nabla_{x} \mathbf{F}_{1}[\mathbf{x}; \{q, T\}]{\bigg \vert}_{\mathbf{x}=\mathbf{1}} \cdot ~\nabla_{x} \mathbf{H}_{1}(x; \{q, T\}){\bigg \vert}_{x=1}~. 
\label{multi10}
\end{equation}
Explicitly, Eq.~\ref{multi10} becomes  
\begin{equation}
{\nabla_{x} \mathbf{H}_{1}(x; \{q, T\})}{\bigg \vert}_{x=1} =  {\left[ I - \nabla_{x} \mathbf{F}_{1}[\mathbf{x}=\mathbf{1}; \{q, T\}] \right]}^{-1} \cdot \mathbf{F}_{1}[\mathbf{1}; \{q, T\}] = {\left[ I - \mathcal{F} \right]}^{-1} \cdot  \mathbf{F}_{1}[\mathbf{1}; \{q, T\}]~,
\label{multi11}
\end{equation}
with $\mathcal{F} = \nabla_{x} \mathbf{F}_{1}[\mathbf{x}=\mathbf{1}; \{q, T\}]$. Introducing the expression in Eq.~\ref{multi8},
we obtain 
\begin{equation}
\langle \mathbf{s} \rangle = \mathbf{F}_{0}[\mathbf{1}; \{q, T\}] + \nabla_{x} \mathbf{F}_{0}[\mathbf{x}=\mathbf{1}; \{q, T\}] \cdot {\left[ I - \mathcal{F} \right]}^{-1} \cdot  \mathbf{F}_{1}[\mathbf{1}; \{q, T\}]~.
\label{multi12}
\end{equation}
The condition for a giant component to emerge consists in the divergence of at least one of the components of $\langle \mathbf{s} \rangle$, that corresponds to the vanishing of the determinant of the matrix $\left[ I- \mathcal{F} \right]$.

\end{document}